\documentclass{aa}

\usepackage[utf8]{inputenc}
\PassOptionsToPackage{hyphens}{url}
\usepackage[english]{babel}
\usepackage{graphicx}
\usepackage{amsmath}
\usepackage[varg]{txfonts}
\usepackage[labelformat=simple]{subcaption}
\usepackage{booktabs}
\usepackage{orcidlink}
\usepackage{tabularx}
\usepackage{ulem}
\usepackage{units}
\usepackage{makecell}
\usepackage{natbib}
\usepackage{hyperref}
\hypersetup{
    colorlinks=true,
    linkcolor=blue,
    citecolor=blue,
    urlcolor=blue,
    breaklinks=true
}

\begin{document}

    \title{The TEQUILA catalog of variables in TESS full-frame images}
   \subtitle{Differential photometry light curves from the first two years of observations}

   \author{Bisi Bernard Ogunwale   
          \inst{1, 2}\fnmsep \thanks{\email{bisi.ogunwale@msmail.ariel.ac.il}}\orcidlink{https://orcid.org/0009-0001-7046-0446}
          \and
          Yossi Zaguri\inst{1, 3}\orcidlink{https://orcid.org/0000-0002-1492-3602}
          \and
          Volker Perdelwitz \inst{4}\orcidlink{https://orcid.org/0000-0002-6859-0882}
          \and
          Marcel V\"olschow \inst{5, 6}\orcidlink{https://orcid.org/0000-0002-2592-2103}
          \and
         Sagi Yosef Azulay\inst{3}
          \and           
          Dafne Guetta\inst{1, 3}\orcidlink{https://orcid.org/0000-0002-7349-1109}
          \and
          Lev Tal-Or\inst{1, 3}\fnmsep \thanks{\email{levtalor@ariel.ac.il}}\orcidlink{https://orcid.org/0000-0003-3757-1440}
          }

   \institute{Department of Physics, Ariel University, Ariel 40700, Israel,
        \and
        Kavli Institute for Astrophysics and Space Research, Massachusetts Institute of Technology, Cambridge, MA 02139, USA
        \and
        Astrophysics, Geophysics, And Space Science Research Center, Ariel University, Ariel 40700, Israel
        \and
        Department of Earth and Planetary Science, 
        Weizmann Institute of Science, Rehovot, Israel
        \and
        Center for X-ray and Nano Science, Deutsches Elektronen-Synchrotron DESY, Germany
        \and
        Department of Information and Electrical Engineering, Hamburg University of Applied Sciences, Germany
             }

   \date{Received: 30 June 2025; Revised: 05 October 2025; Accepted: 05 November 2025}

\abstract
{Stellar variability and transient events provide critical insights into many areas of astrophysics.  Progress in these fields has been accelerated by high-precision space-based photometry missions such as CoRoT, Kepler, and K2. NASA's ongoing Transiting Exoplanet Survey Satellite (TESS) represents another significant milestone, offering a unique combination of long observational baseline, high cadence, and nearly all-sky coverage. However,  extracting high-quality light curves from  TESS full-frame images (FFIs) remains challenging due to contamination from scattered light, primarily from Earth or the Moon, and source blending in crowded fields.}
{In this study, we processed TESS FFIs to produce a comprehensive catalog of light curves for variable point sources observed during the satellite’s prime mission. The resulting database is named TESS quick-look and light curve analysis (TEQUILA) and is intended to support diverse scientific investigations, enable large-scale statistical studies of stellar variability and transient phenomena, and relieve researchers of the need to process TESS FFIs from raw pixel data.}
{We applied the difference image analysis technique, constructing high signal-to-noise photometric reference images via the median combination of quality-filtered FFIs for each charge-coupled device and camera across TESS sectors 1–26. An iterative subtraction method was applied to mitigate instrumental systematics and other variable background features. Light curves were created using simple aperture photometry with a fixed 3-pixel radius centered on sources whose brightness was found to vary significantly in one of the residual images.}
{Our pipeline yields over six million light curves of variable point sources from the first two years of TESS data. These include stellar variables, transient events, instrumental systematics, and moving objects. Approximately \(6 \times 10^{5}\) light curves span multiple sectors, with around \(10^{3}\) originating from the continuous viewing zones. In the median normalized light curves, we achieve a median point-to-point differential variability noise level ranging from \(10^{-3}\) to \(10^{0}\) for sources between 5.0~\(T_{\text{mag}} \) and 16.0~\(T_{\text{mag}}\), while the typical photometric root mean square variability ranges from \(10^{-2}\) to \(10^{1}\). To identify light curves whose creation was prompted by instrumental systematic noise, we employed a convolutional neural network trained in a supervised learning framework. A score was assigned to each classification, reflecting the network’s confidence in the predicted class. To avoid confusion between astrophysical variables and Solar System objects (SSOs), we also include in the catalog a flag that identifies light curves whose creation was prompted by known SSOs.
}
{All extracted light curves are publicly accessible as a high-level science product through the Mikulski Archive for Space Telescopes (MAST). The new catalog can be used as a discovery tool for previously unknown variable point sources, such as astrophysical transients and moving SSOs. In future works, we aim to refine our methods, mitigate remaining systematics, classify the light curves by their phenomenological characteristics, analyze some of the newfound variables, and extend the catalog to include observations from the TESS extended mission.}


\maketitle
%
\section{Introduction}\label{sec:intro}
Variable stars and transient events, the key objects of interest in time-domain astronomy, offer crucial opportunities for investigating various astrophysical phenomena and advancing our understanding of the Universe. These objects exhibit brightness (flux) variation that can be periodic, quasi-periodic, or erratic \citep{2015pust.book.Catelan}, with timescales ranging from thousandths of a second, such as gamma ray bursts, to decades or centuries in luminous blue variables \citep{2008JPhCS.118a2010E}. The cause of such variation can be intrinsic, such as stellar pulsation, or extrinsic factors, such as rotational effect or interaction with other bodies (see Figure 1 of \citet{2008JPhCS.118a2010E} for a summary tree diagram). Their importance cuts across multiple fields of astrophysics, from asteroseismology \citep{2021Aerts}, which analyzes subtle pulsations on the stellar surfaces to probe their interiors, to cosmology \citep{2024Perivolaropoulos}, for example, determining the expansion rate of the local Universe. Additionally, they offer vital testing grounds for developing, verifying, and refining theoretical models \citep{2011Percy}, classifying celestial objects (asteroids, stars, galaxies, etc.) into subcategories, and driving observational studies that lead to the discovery of new bodies, such as exoplanets, comets, and transient events. Advancements in observational technologies and the deployment of sophisticated ground- and space-based observatories have significantly enhanced the study of variable stars and transient events. These developments provide powerful tools for exploring previously inaccessible phenomena, improving our understanding, constraining model parameters, and uncovering new phenomena in this astronomical domain.

\begin{table*}
\centering
\caption{
    Summary of selected TESS FFI light-curve projects, their primary scientific goals, photometric methods, and approximate magnitude limits. 
    \label{tab:ffiprojects}
}
\begin{tabular}{lllll}
\hline\hline
Project Name & Scientific Goal(s) & Photometric Method & Mag. Limit (TESS mag) & References \\
\midrule
MIT QLP        & Transiting exoplanets         & Simple Aperture Photometry & $\lesssim 13.5$  & (1) \\
TESS-SPOC      & Transiting exoplanets         & Simple Aperture Photometry & $\lesssim 16$  & (2) \\
TGLC           & Transiting exoplanets         & Aperture + PSF Photometry  & $\lesssim 16$  & (3) \\
GSFC-ELEANOR   & Transit / stellar variability & Aperture Photometry        & $\lesssim 16$  & (4) \\
CDIPS          & Transiting exoplanets         & Image Subtraction          & $\lesssim 16$  & (5) \\
PATHOS         & Transiting exoplanets         & PSF Photometry             & $\lesssim 16.5$  & (6) \\
DIAMANTE & Transiting exoplanets & Image Subtraction & 
\makecell[l]{FGK: $V \leq 13$ \\ M-dwarfs: $V \leq 16$, $d < 600$ pc} & (7)(8) \\
TASOC          & Asteroseismology              & Simple Aperture Photometry & $\lesssim 15$  & (9) \\
T16            & Transit / stellar variability & Image Subtraction          & $\lesssim 16$  & (10) \\
TEQUILA & All variable sources & Image Subtraction & No imposed limit & (11) \\

\bottomrule
\end{tabular}
\tablebib{
(1) \citet{2020Huang}; 
(2) \citet{2020Caldwell}; 
(3) \citet{2023Han}; 
(4) \citet{2022Powel}; 
(5) \citet{2019Bouma}; 
(6) \citet{2019Nardiello}; 
(7) \citet{2020Montalto}; 
(8) \citet{2023Montalto}; 
(9) \citet{2021Handberg}; 
(10) \citet{2025Hartman_T16}.
(11) This work.
}
\end{table*}

\begin{table*}
\centering
\caption{
    Overview of notable tools for processing TESS FFIs.
    \label{tab:tess_tools}
}
\begin{tabular}{lll}
\hline\hline
Tool & Description & References\\
\hline
\texttt{Lightkurve} & light curve extraction from TPFs, cutouts, and other analyses. & (1) \\
\texttt{TESSCut} & Extraction tool for TPF-like cutouts from TESS FFIs & (2) \\
\texttt{eleanor} & Stellar light curve extraction pipeline & (3) \\
\texttt{TESSreduce} & Transient light curve extraction tool for TESS FFIs. & (4) \\
DIA & Differenced images light curve extraction software from FFIs& (5) \\
ISIS &  Differenced images light curve extraction pipeline for TESS FFIs cutout.& (6)\\
\texttt{TESSELLATE} & Untargeted sector-wide image subtraction and sources detection based on \texttt{TESSreducde} & (7)  \\
\hline
\end{tabular}
\tablebib{
 (1) \citet{2018ascl.soft12013L}; (2) \citet{2019Brasseur}; (3) \citet{2019Feinstein}; (4) \citet{2021Ryan_reduce}; 
(5) \citet{2018Oelkers}; (6) \citet{2021Fausnaugh} ; (7) \citet{2025Roxburgh}
}
\end{table*}

The \textit{Transiting Exoplanet Survey Satellite} \citep[TESS, ][]{2015Ricker} is an ongoing NASA mission carrying out a near-all-sky survey. Launched in April 2018 with the primary goal of detecting and characterizing exoplanets around bright stars in the Solar neighborhood, the mission has significantly contributed to time-domain astronomy. The satellite is equipped with four 10cm telescopes, with each one featuring four charge-coupled device (CCD) cameras covering 24\textdegree $\times$ 96\textdegree patches of the sky known as a sector, observed for $\sim$27 days. TESS data products are in two major categories: the short-cadence target pixel files (TPFs), also known as  ``postage stamps'' of $\sim$10$^5$ preselected targets per sector, and the long-cadence full frame images (FFIs) of the entire field of view (FOV). The TPFs were produced at a 120-second cadence during the survey's prime mission (between 25 July 2018 and 4 July 2020), while the FFIs were generated in 1800 seconds \citep{2018Vanderspek}\footnote{\url{https://archive.stsci.edu/missions/tess/doc/TESS_Instrument_Handbook_v0.1.pdf}}. In the first extended mission, an additional 20-second cadence of TPFs was created, while the cadence of the FFIs was increased to 600 seconds. The FFI cadence was further increased to 200 seconds in the second extended mission and onward. TPFs are processed into light curves at the mission's Science Processing Operations Center (SPOC) \citep{2016Jenkins}, which is optimized for planetary transit search by correcting instrumental systematics and stellar long-term trends that might obscure the transit signals. The SPOC photometric pipeline performs simple aperture photometry (SAP) and presearch data conditioning (PDC)-SAP light curves, representing the raw and co-trended corrected flux time series. 

The TESS, with its high-precision photometry data and the large FOV, has become an invaluable asset in the time-domain astronomy community. As of April 2025, the mission has led to the discovery of 620 confirmed exoplanets and 7,576 candidates\footnote{Current statistics available at \url{https://exoplanetarchive.ipac.caltech.edu/docs/counts_detail.html}}. While TESS was primarily designed for detecting transiting exoplanets, much of its broader scientific impact stems from the FFIs. These data products, though demanding sophisticated reduction techniques, have enabled a wide range of time-domain studies. Examples include the investigation of binary stars \citep{2022Prsa}, tidal disruption events \citep{20191Holoien}, active galactic nuclei \citep{2023Treiber}, RR Lyrae stars \citep{2022Molnar}, Solar System bodies \citep{2025Takacs,McNeill_2023,2020Pal}, stellar pulsations and asteroseismology \citep{2024Hey}, supernovae \citep{2021Fausnaugh}, and gamma-ray burst afterglows \citep{2024Roxburgh, 2024Jayaraman}. For a comprehensive overview of TESS’s mission history, operational strategy, exoplanetary discoveries, and broader scientific contributions, see the recent review by \cite{2024Winn}.

While timeseries (light curve) data are readily available for the preselected stars in the short-cadence data, the reduction of FFIs necessitates significant efforts from researchers, typically producing high-level science products (HLSPs) most accessible via MAST\footnote{HLSP available at \url{https://mast.stsci.edu/hlsp/\#/}}.
 However, these projects are predominantly focused on specific scientific goals, mainly exoplanet discovery, which restricts their broader applicability to other astrophysical research areas. For instance, the MIT quick look pipeline (QLP) \citep{Kunimoto_2021, 2020Huang} generates light curves for stars up to a TESS magnitude ($T_{\text{mag}}$) of 13.5 using the TESS input catalog \citep[TIC,][]{Stassun_2018, Stassun_2019}. The pipeline relies on FFIs calibrated by TESS image calibration (TICA) \citep{2020Fausnaugh_tica}, and the resulting light curves are detrended using a high-pass filter to emphasize transit signals over stellar variability. While this approach enhances transit detection, it significantly hampers transient phenomena and stellar variability studies, where long-term trend signals are critical.

In addition to the MIT-QLP light curves, the community has executed several TESS FFI light curve projects. One notable HLSP is the TESS data for asteroseismology (T'DA) \citep{2021Handberg, 2021Lund, 2021Audenaert}, developed by the TESS asteroseismology consortium (TASOC) \citep{2017Lund} to support research in asteroseismology. \citet{2023Han} introduced PSF-based TESS FFI light curves for stars as faint as $T_{\text{mag}}$ 16, using Gaia DR3 astrometry and the targets' magnitudes as priors. 

Additional HLSP light curves include the TESS-SPOC light curve \citep{2020Caldwell}, which focuses on a selected subset of stars from the TIC. Other HLSPs from TESS FFIs include the NASA GSFC TESS FFI light curves \citep{2022Powel}, the Cluster difference image photometric survey (CDIPS) \citep{2019Bouma}, PATHOS \citep{2019Nardiello}—a PSF-based survey of stellar clusters—and DIAmante light curves \citep{2020Montalto}. Recently, \citet{2025Hartman_T16} generated light curves for stars within the TESS FOV, reaching up to $T_{\text{mag}}$ 16 during the satellite's first year of operation.

In addition to the timeseries data HSLP, the TESS community has developed several tools for generating custom light curves from the mission's FFIs. These tools enable users to extract light curves from specific regions of interest within the FFIs and perform various processing tasks. For instance, \texttt{tesscut} \citep{2019Brasseur} allows users to select and extract small pixels, akin to TPF, from a specified CCD, camera, and sector. The resulting data cube can then be used for further analysis. Another widely used tool is \texttt{lightkurve} \citep{2018ascl.soft12013L}, a Python package, which supports a broader range of astronomical data analyses for TESS and Kepler. 

Other notable TESS tools include the \texttt{eleanor} package \citep{2019Feinstein}, which can create light curves optimized for transit searches from FFIs. It first creates a ``postcard'' of 148 × 104 pixels, where background systematics are determined and removed. This postcard creates target TPFs of 13 × 13 pixels centered on the target. Users can select aperture or PSF photometry depending on their analysis needs.\texttt{TESSreduce} \citep{2021Ryan_reduce} is another key tool for producing light curves from TESS FFIs, particularly for transient events. Because such signals are often biased by scattered light, \texttt{TESSreduce} carefully models the background by accounting for detector stripe artifacts and masking known sources within the cutout region to avoid overestimation, before determining the final background model. Other tools for extracting light curves from TESS FFIs include those based on difference image analysis (DIA) \citep{2008Miller,  2000Alard, 1998Alard_Lupton}, such as the ISIS-based pipelines, for example, \citet{2019Valley, 2021Fausnaugh} and the DIA pipeline for TESS\footnote{\url{https://github.com/ryanoelkers/DIA}} \citep{2018Oelkers}. While preparing this manuscript, \citet{2025Roxburgh} released \texttt{TESSELATE}, a pipeline built upon \texttt{TESSreduce} to perform a sector-wide variability search in TESS FFIs. A summary of various publicly available software that has been developed for processing TESS FFIs is provided in Table \ref{tab:tess_tools}. 

Except for \texttt{TESSELATE}, which allows for an untargeted search for variable sources in TESS FFI, most of the tools developed to date for processing TESS FFIs require some level of user input. A significant limitation of these tools is that users must manually enter relevant information for each target of interest. This manual input can become impractical when dealing with many objects within the field of view. This bottleneck presents a significant challenge, especially for large-scale studies requiring efficient and automated light curve extraction for numerous sources.

In this research project, we undertook a comprehensive extraction of all variable point sources across the FFIs, utilizing DIA. This approach allows us to extract light curves for both spatially varying point sources, such as moving Solar System objects (e.g., asteroids), and temporally varying point sources, including variable stars and transient events, across the entire TESS field of view during its first two years of scientific operation. To our knowledge, this represents the first large-scale catalog of variable point sources created from TESS FFIs. 

We outline our methodology for the light curve extraction using DIA in Sec. \ref{sec:method}. In Sec. \ref{sec_result}, we discuss our results and give descriptive statistics of the light curves, including the photometric precision and noise levels. Sec. \ref{sec:productdesc} describes the catalog, its products, and data access on MAST. Finally, Sec.~\ref{sec:discussion} discusses our findings, and Sec.~\ref{sec:future_plans} outlines future directions and planned developments.

\section{Methods}\label{sec:method}

This section outlines the process of extracting light curves using the DIA technique \citep{2008Miller, 2000Alard, 1998Alard_Lupton}. Before the launch of TESS, \cite {2018Oelkers} demonstrated the effectiveness of this method in extracting light curves from simulated FFIs, and it has been applied in different TESS light curve extraction projects with slight variations. While the work presented here builds on the general methodology outlined by \citet{2018Oelkers}, the TEQUILA framework was developed independently and differs in several key aspects, including the background determination and other components of the formalism. The framework was first introduced in \citet{thomfohrde2025algorithm}, and in this paper, we extend it to all available TESS FFIs from the first two years of the mission to construct a comprehensive catalog of candidate variable sources.

\begin{figure}
   \centering
   \includegraphics[width=0.5\textwidth, height=0.5\textheight, keepaspectratio]{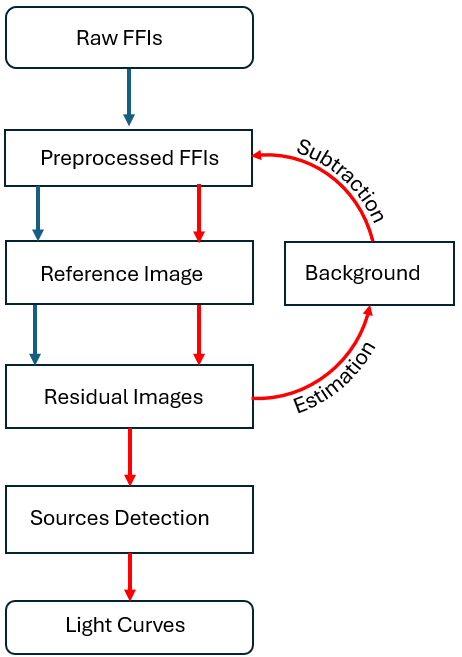}
   \caption{Schematic diagram showing our step-by-step photometry processing method.}
 \label{Fig_Method}%
    \end{figure}

We retrieved the calibrated FFIs from the Mikulski Archive for Space Telescopes (MAST)\footnote{\url{https://archive.stsci.edu/tess/bulk_downloads/bulk_downloads_ffi-tp-lc-dv.html}}. For each CCD, camera, and sector combination, our procedure involved: (i) preprocessing each image, (ii) constructing a photometric reference image via median combination of all the images, (iii) modeling the background for each image, (iv) subtracting each image from its corresponding CCD-camera-sector reference, (v) identifying sources in the residual images, (vi) calculating the flux to generate light curves for the detected sources from each residual images. A detailed flowchart of this extraction process is provided in Fig. \ref{Fig_Method}, and the pipeline is publicly available on Github\footnote{\url{https://github.com/Astro-informatics-cipher-project/image-subtraction}}.

\subsection{Image preprocessing}

The extraction process began with a series of preprocessing steps applied to all the FFIs in a given CCD. This included iterating through the dataset to assess the image quality and extract metadata from the FITS headers. FFIs that lacked critical information, such as valid World Coordinate System (WCS) data or were flagged with data quality issues, were excluded from further analysis.

We trimmed the FFIs at this stage by removing the virtual rows and columns using the \texttt{Cutout2D} function from the \texttt{astropy.nddata} module. The original TESS FFIs, which include these virtual rows and columns, have dimensions of \(2078 \times 2136\), while the actual science data has dimensions of \(2048 \times 2048\).

\subsection{Image alignment}
Accurate image alignment is crucial for reliable image subtraction. Although the TESS pointing is generally stable within sub-pixel levels, minor pixel-to-pixel variations between images can introduce trends into the resulting light curves. To mitigate this, we align all images using the WCS information contained in the image headers. For simplicity, we selected the first image in the datasets of a given CCD as the astrometric reference image and aligned the subsequent images to it. The alignment is performed using the \texttt{hcongrid} class from the \texttt{FITS\_tools} package.
    
\subsection{Reference image}
The photometric reference image, sometimes referred to as a master frame, is a crucial component of DIA. This reference image is typically characterized by a high signal-to-noise ratio (S/N) and, in ground-based observations, is selected based on the best seeing conditions \citep[e.g.,][]{1998Alard_Lupton}, which strongly affect the PSF of images from one image to another. Although atmospheric seeing is not a concern for space-based observatories such as TESS, the mission's pixel response function (PRF) exhibits significant spatial variation across the FFI due to the telescope's wide FOV, as well as temporal variation from image to image, which must be accounted for. 

Various methods have been employed to create or select reference images in TESS FFI subtractions. Some approaches involve selecting a single image from the dataset \citep[e.g.,][]{2019Oelkers}. In contrast, others utilize median stacking of images over specific epoch intervals \citep[e.g.,][]{2019Valley} or median stacking of a subset of images based on particular criteria \citep[e.g.,][]{2021Fausnaugh}.

In our pipeline, we adopted the median combination technique outlined by \citet{2018Oelkers}. Specifically, we median-stack all images in a CCD in batches of 50, and the resulting stacked images (intermediate master frames) are further median-stacked to produce the final reference image. Although this method may be computationally expensive, it ensures the highest possible S/N for the reference image.

\subsection{Background estimation and removal}

Background estimation is among the most challenging tasks when working with TESS FFIs due to the presence of complex, large-scale background variation caused primarily by scattered light from Earth and the Moon. Various TESS FFIs reduction software packages for light curve extraction have adopted different techniques to address this issue. Many of these approaches rely on cutouts of the FFIs to simplify background estimation. However, as discussed in Sec. \ref{sec:intro}, this strategy becomes inefficient for large-scale data reduction such as ours.

TESS FFIs are affected by both intrinsic system factors and external background sources. Intrinsic sources include reflective and conductive straps located at the back of the CCD, while external sources primarily consist of temporally and spatially scattered light from Earth and the Moon.

To address these challenges, we implemented an iterative background estimation approach. Unlike the method described by \citet{2018Oelkers}, which performs background subtraction before the reference image subtraction — a step that may bias the background estimation as noted by \citet{2021Ryan_reduce} - we instead begin by subtracting the reference image from each FFI. This removes constant sources from the frame, leaving behind only the spatially varying background and any residual signals from variable sources. 

We then estimate the background from the residual images using the \texttt{MedianBackground} function from the \texttt{photutils} package \citep{Bradley2024}, with a mesh size of $32 \times 32$ pixels and a median filter of $3 \times 3$. This estimated background is iteratively subtracted from the original FFIs. Following this, we construct a new reference image using the background-subtracted FFIs and repeat the subtraction process. After two iterations, we find that the majority of the background signal has been effectively removed, and further iterations yield negligible improvement.

\subsection{Image subtraction}
Our image subtraction routine follows the methodology described in \citet{2018Oelkers}, incorporating the mathematical framework detailed by \citet{2008Miller}. In brief, we adopt a Dirac \(\delta\)-function kernel, which is particularly well-suited for handling irregular PSF shapes. This is especially important for TESS data, where the significant FOV results in a highly variable PRF across the CCDs. We employ a \(5 \times 5\) pixel convolution kernel and use stamps centered on 500 bright, unsaturated, isolated stars to derive the best-fit kernel coefficients. These coefficients are determined using a least-squares minimization approach. This technique ensures a precise match between the master frames and science images in terms of PSF shape and flux distribution, thereby improving the quality of the residuals and the reliability of subsequent photometric measurements.

\subsection{Source detection}
At this stage, we iterated over the residual images and applied a source detection algorithm to identify potential variable sources in each frame. For source detection, we used the DAOphot star-finding algorithm, implemented as \texttt{DAOStarFinder} in the \texttt{photutils} Python library \citep{Bradley2024}. This routine identifies point sources by locating local maxima in the residual images after background statistics have been estimated. The background level and noise were determined using iterative sigma clipping with $\sigma = 3.0$ and a maximum of five iterations. This ensures that bright stars, cosmic rays, and other outliers are excluded from the background calculation. For the star-finding parameters, we adopted a full-width at half maximum (FWHM) of 1.5 pixels, and the detection threshold was set to 5.0 times the background RMS, a deliberately conservative choice to minimize spurious detections from noise. To avoid edge artifacts, we excluded all detections within three pixels of the image boundary, consistent with the photometric aperture used later.

In a residual frame, sources that are brighter in the FFI than in the reference image appear as positive peaks, whereas those that are fainter (e.g., during eclipses) appear as negative features. To detect both cases, we applied the detection algorithm to each residual image and to its inverted counterpart, ensuring that negative sources were also identified. We did not take the absolute value of the residual images, since inversion preserves the spatial profile of fainter sources more accurately. Despite the fact that TESS PSF is slightly undersampled in the central parts of each camera's optical axis \citep{2018Vanderspek}, we did not mask bright targets, even though they might have poor subtractions.

To determine whether detections across the sequence of residual images correspond to the same astrophysical source, we cross-matched the source lists against a reference list constructed from the first residual image in each CCD. A new detection not present in the reference list was added, while duplicates were ignored. Sources were considered the same if they lay within 3 pixels of one another. This procedure produced a comprehensive table of all candidate sources for each CCD.

   \begin{figure}
   \centering
   \includegraphics[width=0.49\textwidth, height=0.6\textheight, keepaspectratio]{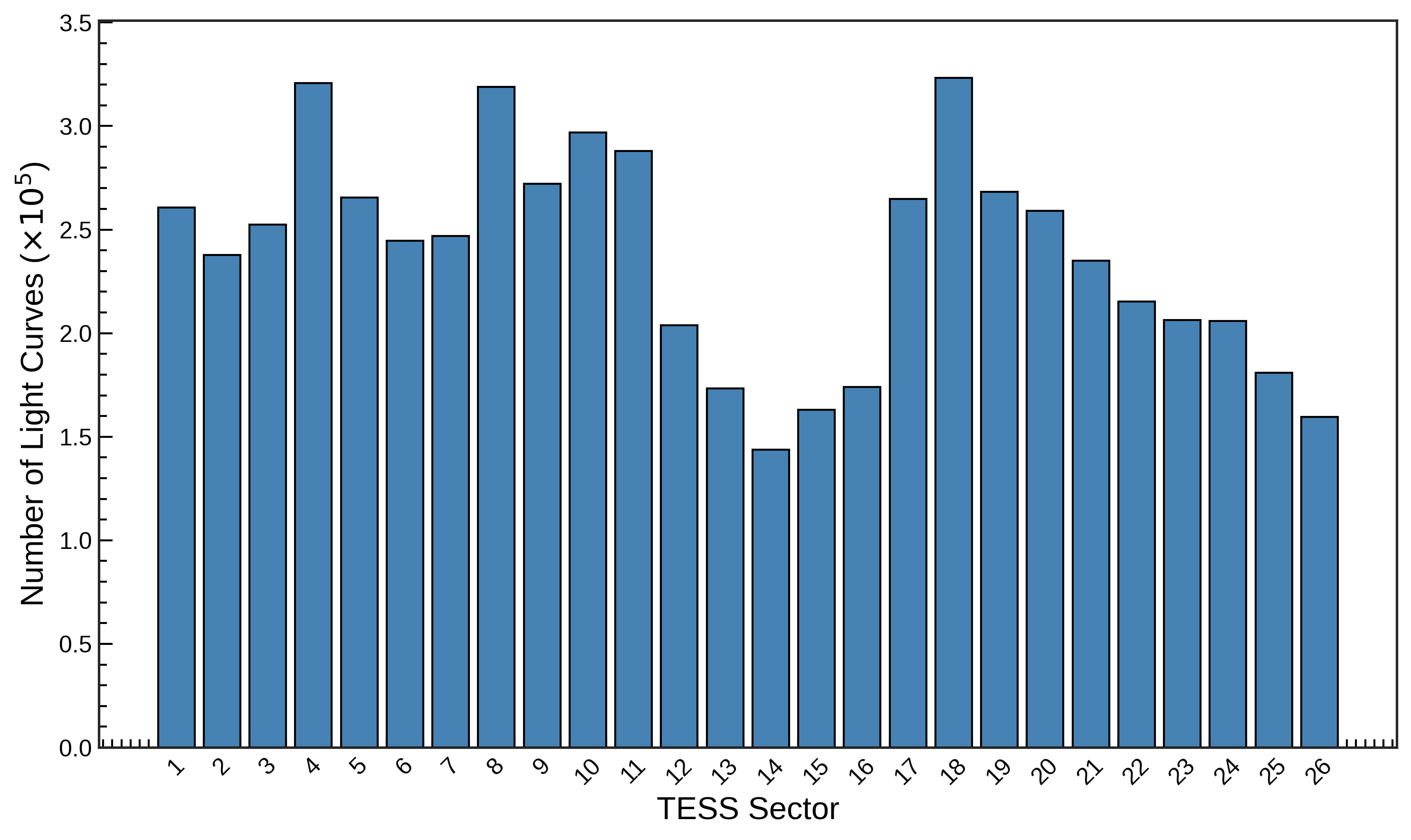}
   \caption{Distribution of variable sources light curves extracted in each sector of the TESS prime mission.}
              \label{Fig_Num_LC}%
    \end{figure}

\subsection{Light curve extraction}
The light curves of the identified variable sources are extracted using aperture photometry with a fixed 3-pixel radius centered on each source on the residual images. This results in the differential flux time-series labeled \texttt{d\_FLUX} and their estimated differential flux error (\texttt{d\_FLUX\_ERR}). The resulting light curves are saved as binary \texttt{fits} files compiled using the \texttt{Lightkurve} Python package \citep{2018ascl.soft12013L}.


   \begin{figure*}
   \centering
   \includegraphics[width=0.48\textwidth]{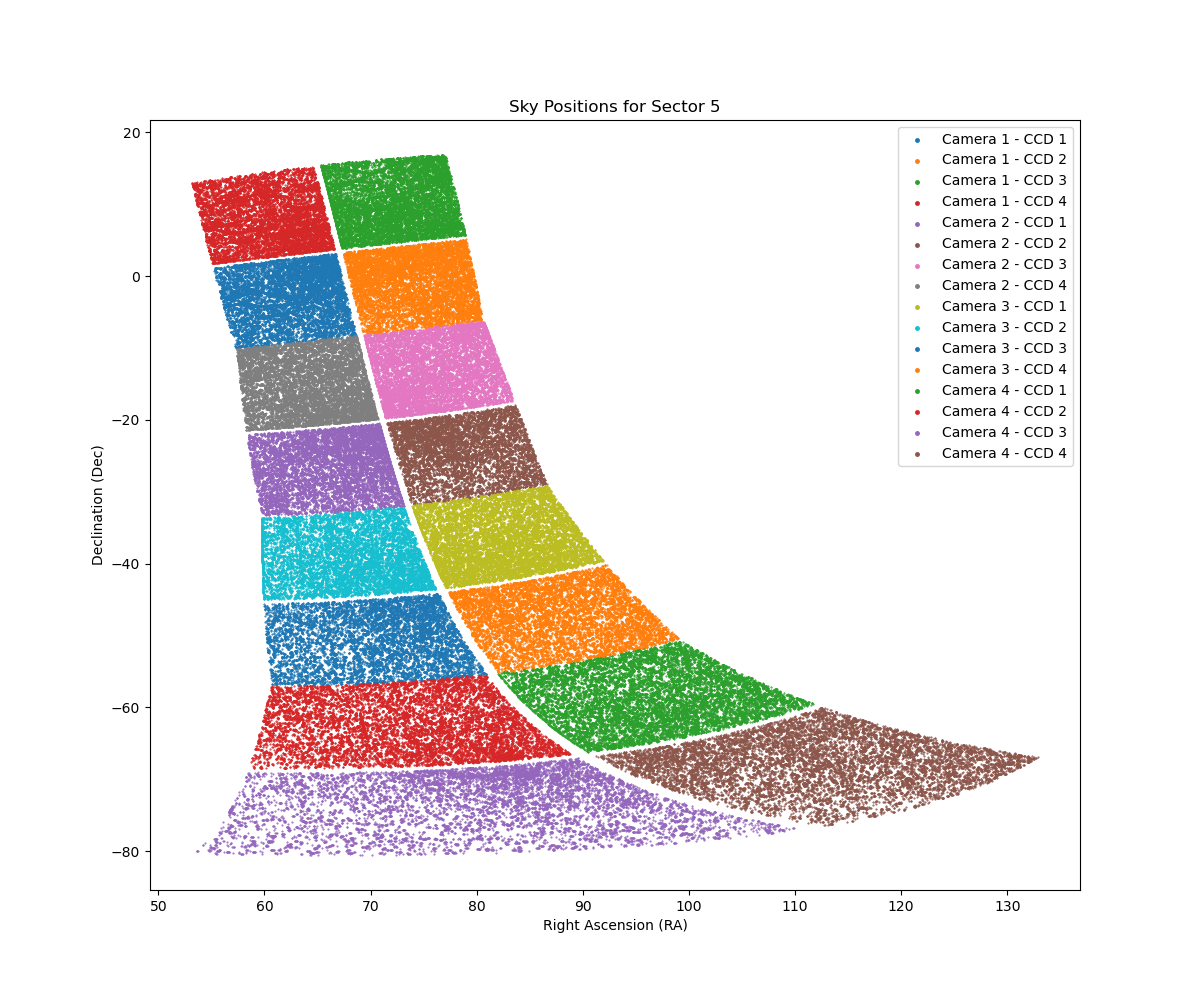}
    \includegraphics[width=0.48\textwidth]{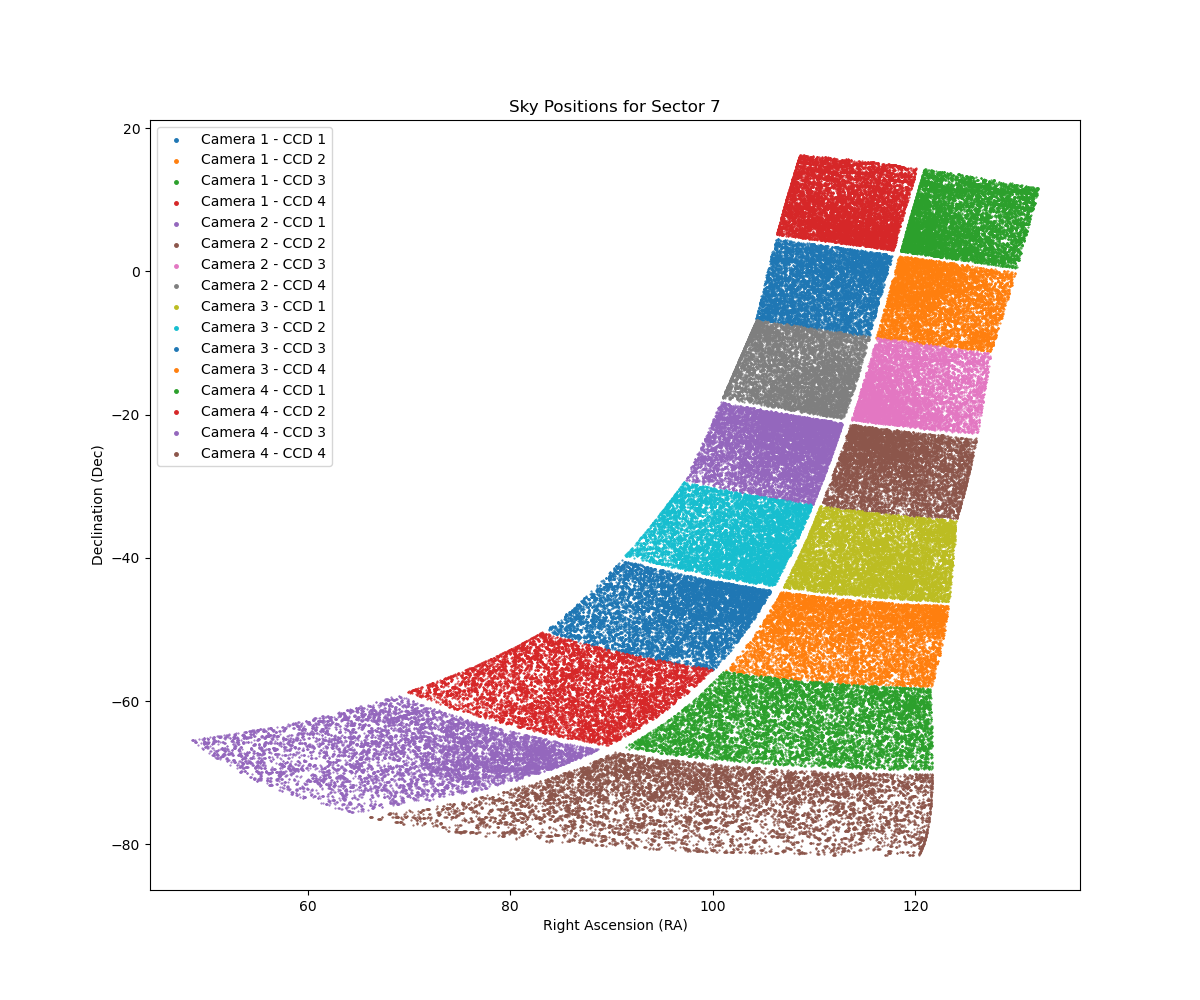}
   \caption{On-sky plots of the coordinates for the extracted variable point sources in sectors 5 and 7.}
              \label{Fig_On_sky_plot}%
    \end{figure*}

\begin{figure*}
   \centering
   \begin{subfigure}{0.48\textwidth}
       \includegraphics[width=\textwidth]{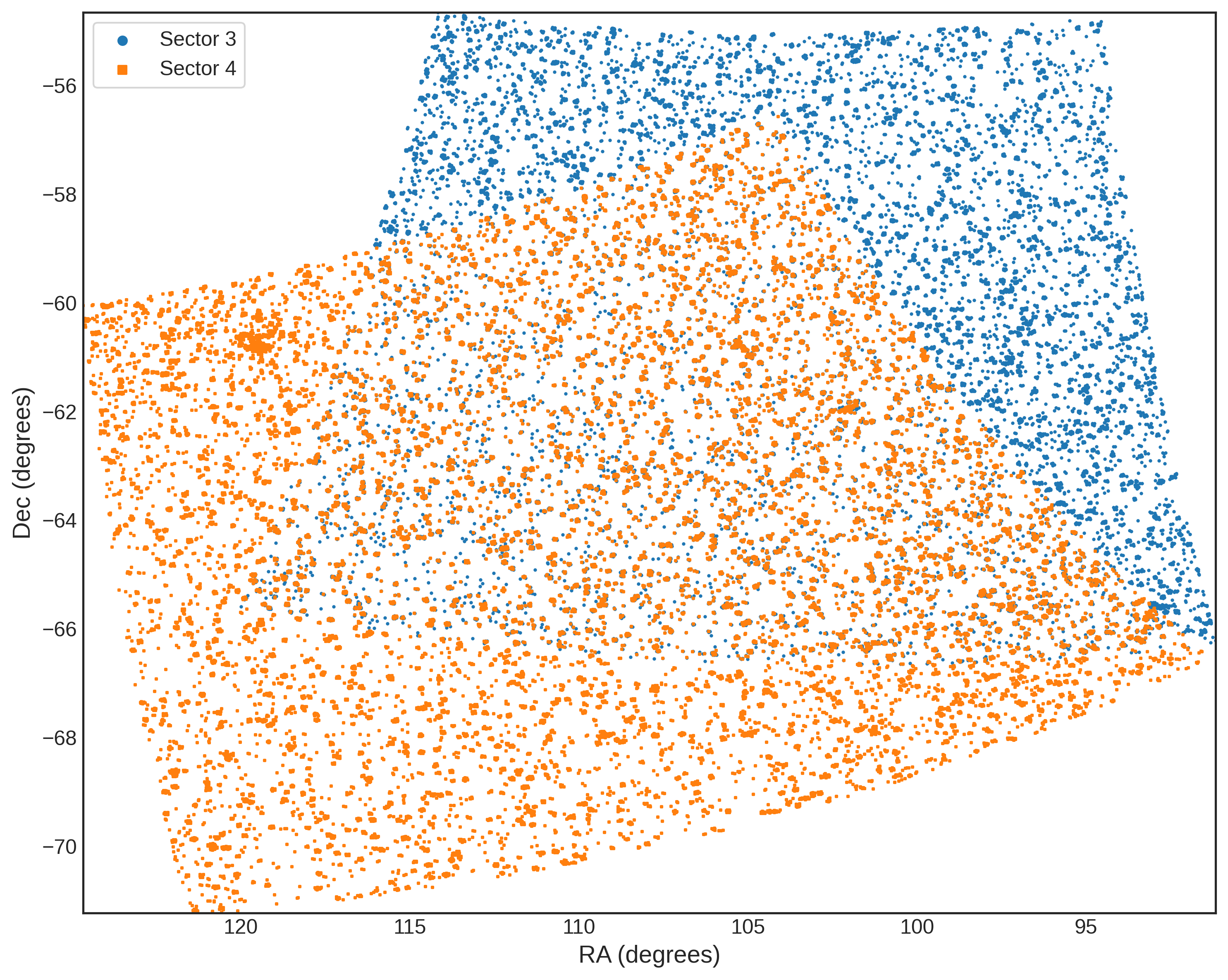}
       \caption{Overlap of extracted sources in camera 4, CCD 4 of sectors 3 \& 4.}
       \label{Fig_overlap}
   \end{subfigure}
   \begin{subfigure}{0.48\textwidth}
       \includegraphics[width=\textwidth]{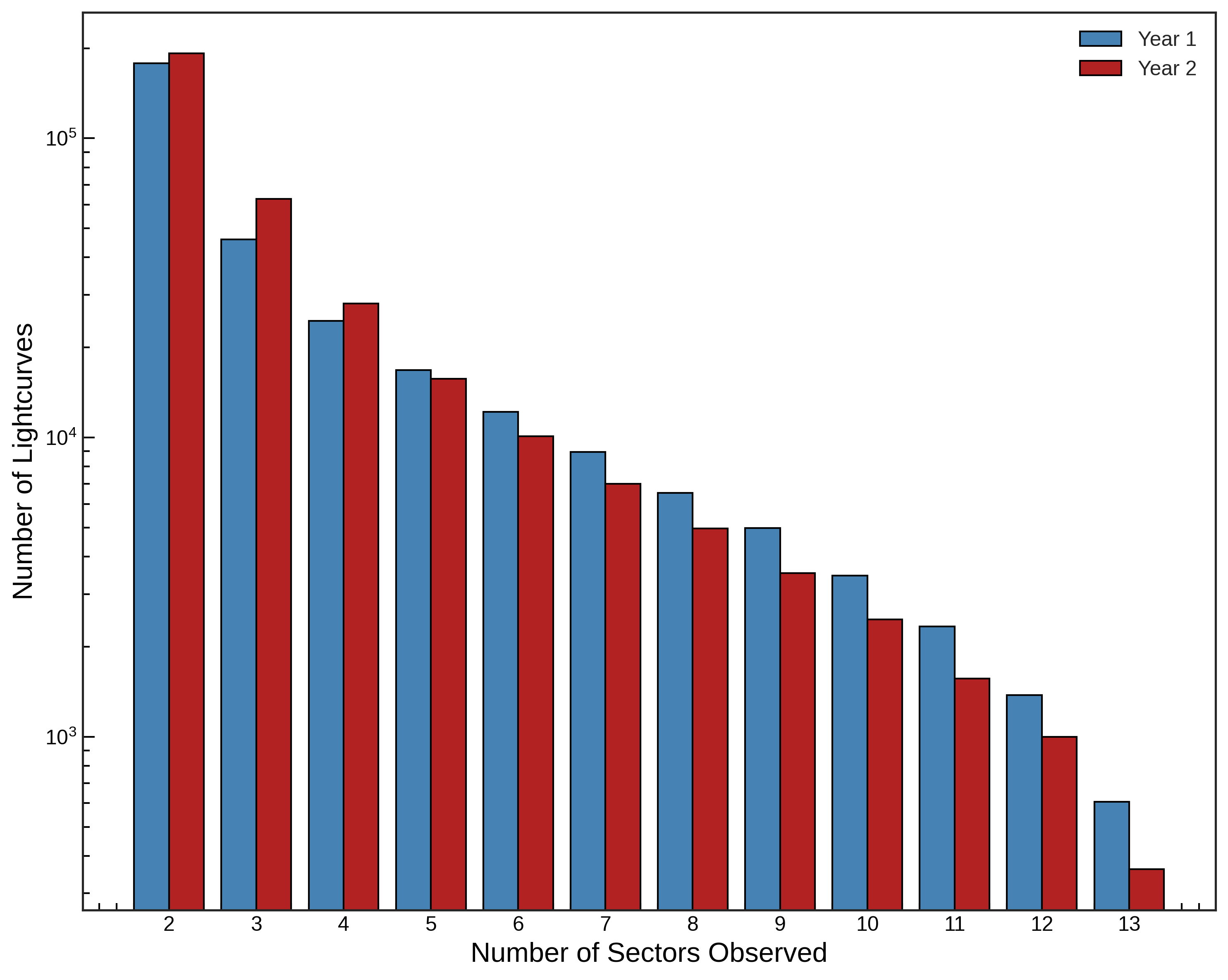}
       \caption{Distribution of multisector sources across Years 1 and 2.}
       \label{Fig_multisector}
   \end{subfigure}
   \caption{Extracted sources with multisector observation and their distributions for the first two years of TESS observation.}
   \label{Fig_result}
\end{figure*}

\subsection{Flux calibration}\label{sub-sec:flux_cal}
To determine the reference flux (\texttt{REF\_FLUX}), we performed forced aperture photometry on the photometric reference frame of each CCD in a camera of a sector for every detected variable source, including their corresponding uncertainties (\texttt{REF\_FLUX\_ERR}). The \texttt{REF\_FLUX} is then added to the \texttt{d\_FLUX} to obtain the flux (\texttt{FLUX}) value at each epoch. The flux uncertainties \texttt{FLUX\_ERR } were estimated as the quadrature sum of the \texttt{d\_FLUX\_ERR} and the median reference flux uncertainty (REF\_FLUX\_ERR). Due to anomalies in some reference images, the reference flux could not be determined for approximately 5400 light curves, and these values are therefore recorded as NaN.

Since we conducted an untargeted variable source search without cross-matching the detected sources with the TIC, we estimated the TESS-equivalent magnitude of each source using the relation  
\begin{equation} \label{eqn:mag_tess}  
    T_{\text{mag}} \approx  T_{0} - 2.5\log_{10}(\text{\texttt{R\_F}}),  
\end{equation}  
where \( T_{0} \) is the magnitude zero-point, and \texttt{R\_F} is the (\texttt{REF\_FLUX}) obtained from forced aperture photometry on the photometric reference images. Following the estimation by \citet{2018Vanderspek}, that for a star of magnitude 10, TESS will yield 15,000 \(e^{-} / s\), we obtained a zero-point value of \( T_{0} = 20.44 \). We note that this estimated magnitude may be overestimated and could differ from the values in the TIC for the stellar source. However, since our main focus is to provide light curves in terms of the flux values, the magnitude was only used to demonstrate the signal and noise levels of the extracted light curves in Sec. \ref{sub_sec:noise_metric}.

\section{Post processing and results}\label{sec_result}

We extracted over 6.1 million variable point sources from the TESS prime mission. The distribution of light curves across each sector is illustrated in Fig. \ref{Fig_Num_LC}. Samples of on-sky plots of the coordinates of extracted variable point sources in sectors 5 and 7 are also presented in Fig. \ref{Fig_On_sky_plot}.

\subsection{Multi-sector light curves}
Sources, mostly stars, located near the ecliptic poles are observed across multiple TESS sectors, with longer observational time baselines depending on their proximity to the ecliptic poles.
In Fig. \ref{Fig_overlap}, we present an example of the overlapping coordinates of sources extracted from camera 4 in sectors 3 and 4. A total of 635,911 light curves from multisector variable source objects were extracted, representing $\sim10$\% of the overall number of light curves. This includes about 1,000 sources located in the continuous viewing zones. Fig. \ref{Fig_multisector} illustrates the distribution of the extracted multisector light curves across Years 1 and 2.

\begin{figure*}
\centering
\includegraphics[width=0.99\textwidth, keepaspectratio]{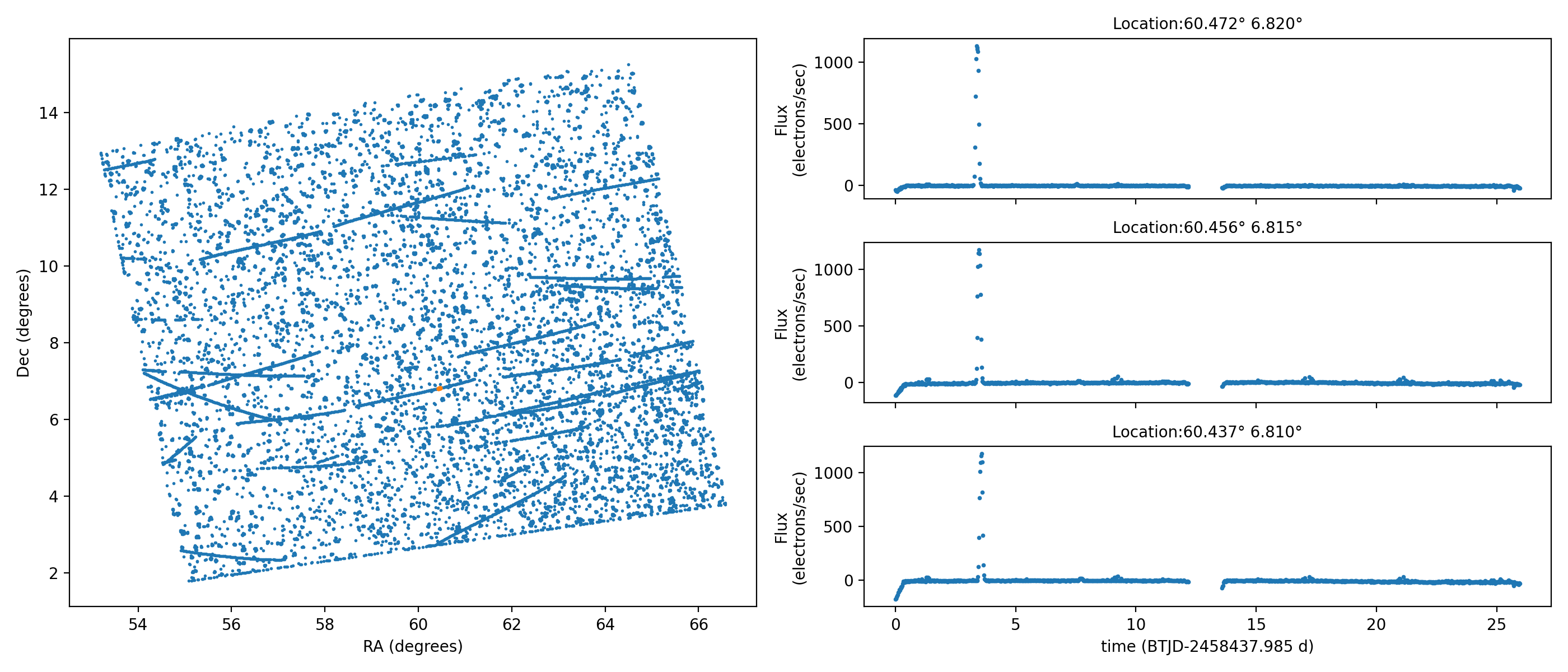}
\caption{Left: On-sky plot of the extracted sources for CCD 4 of Camera 1 of sector 5. The streaks represent paths of Solar System objects as they cross the telescope's field of view during the observation. Orange highlights the location of the three light curves in the right panel. Right: Three light curves whose extraction was prompted by the asteroid 339 Dorothea (A892 SC) crossing the FOV at three consecutive epochs. The spikes represent the increase in flux due to the asteroid crossing the on-sky location of the light curves, where otherwise there is no light source.}
\label{Fig_SSO}%
\end{figure*}
    
\subsection{Solar System objects}
\label{sec:ssos}

Since our pipeline extracts a light curve wherever it detects flux variability, we also detected bright Solar System objects that crossed the TESS FOV during its observation. A sample on-sky plot of the extracted source coordinates for all CCD 4 of Camera 1 in sector 5 is shown in the left panel of Fig. \ref{Fig_SSO}. The distinct streak lines in the plot indicate the paths of Solar System objects. In the right panel of Fig. \ref{Fig_SSO}, we provide an example of three light curves whose extraction was prompted by the asteroid 339 Dorothea (A892 SC) crossing the FOV at different epochs in sector 5.

\begin{figure}
\includegraphics[width=0.49\textwidth, keepaspectratio]{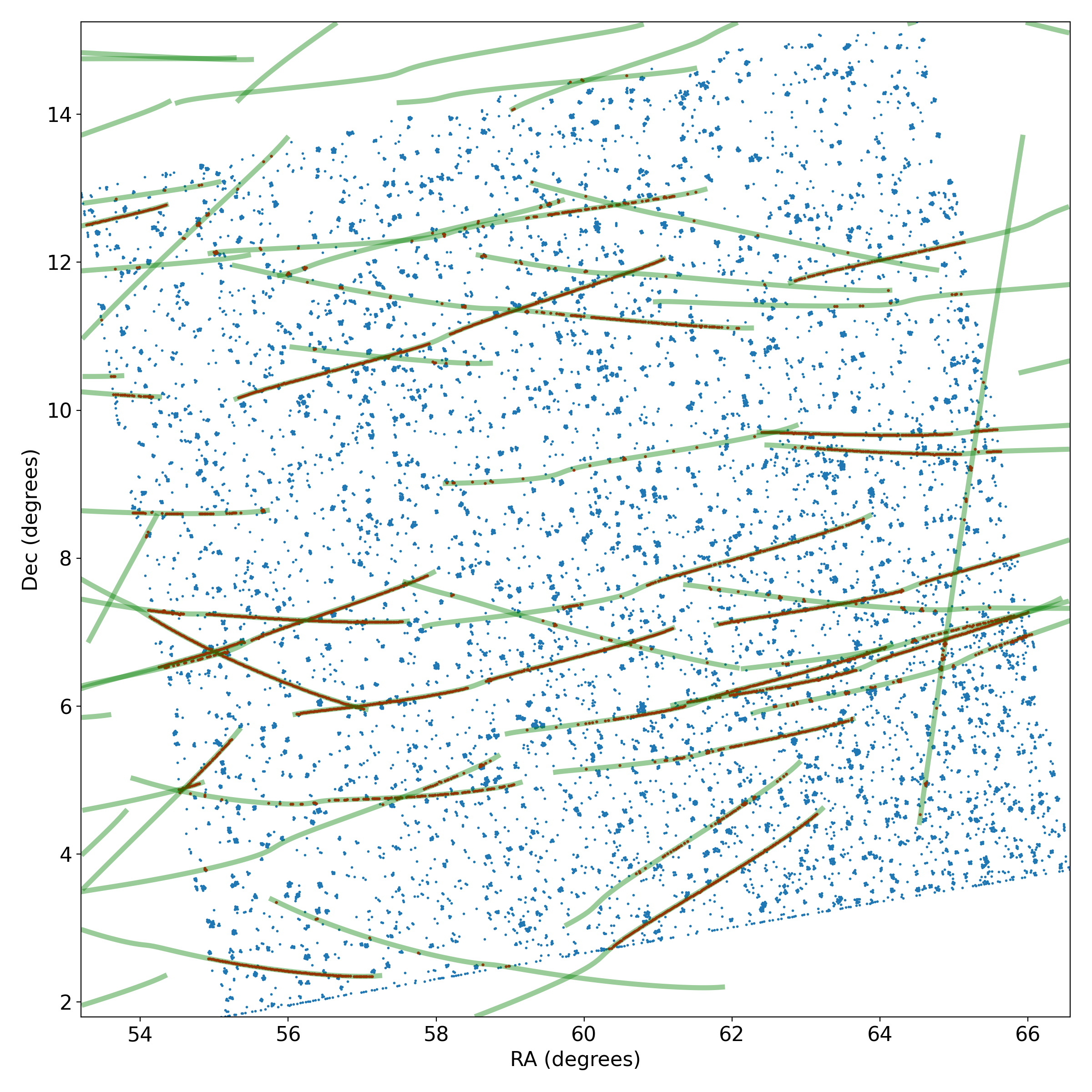}
\caption{Extracted sources for CCD 4 of Camera 1 of sector 5 (blue dots). Green lines represent known SSO trajectories found using the Small-Body Identification Tool. Sources that are located less than one arcminute of such a trajectory are highlighted using the color red and receive the SSO contamination flag.}
\label{Fig_SSO_flag}%
\end{figure}

To avoid confusion between astrophysical variables and SSOs, we include in the catalog a flag that identifies light curves whose creation was prompted by known Solar System objects. To create the flag, we made use of the small-body identification tool\footnote{The NASA JPL small-body identification tool is available at \url{https://ssd.jpl.nasa.gov/tools/sb_ident.html##/)}} via the Python API with a fiducial limiting magnitude of 16.5. Every sector has been queried at 100 uniformly spaced Julian Dates to create a list of SSOs and their positions. Then, we went through all light curves and checked if they lie within one arcminute of an SSO trajectory, employing linear interpolation between existing timestamps to fill the gaps (see Fig.~\ref{Fig_SSO_flag}). Overall, we found $165,678$ light curves whose creation was prompted by known SSOs, corresponding to $2.7$\,\% of the total number of light curves in the current catalog. Since during the nominal TESS mission, camera 1 was always pointing toward low ecliptic latitudes and camera 4 was pointing toward the ecliptic poles, the fraction of SSO-related light curves falls off with camera number.

 \begin{figure*}
    \centering
    \includegraphics[width=1.0\linewidth]{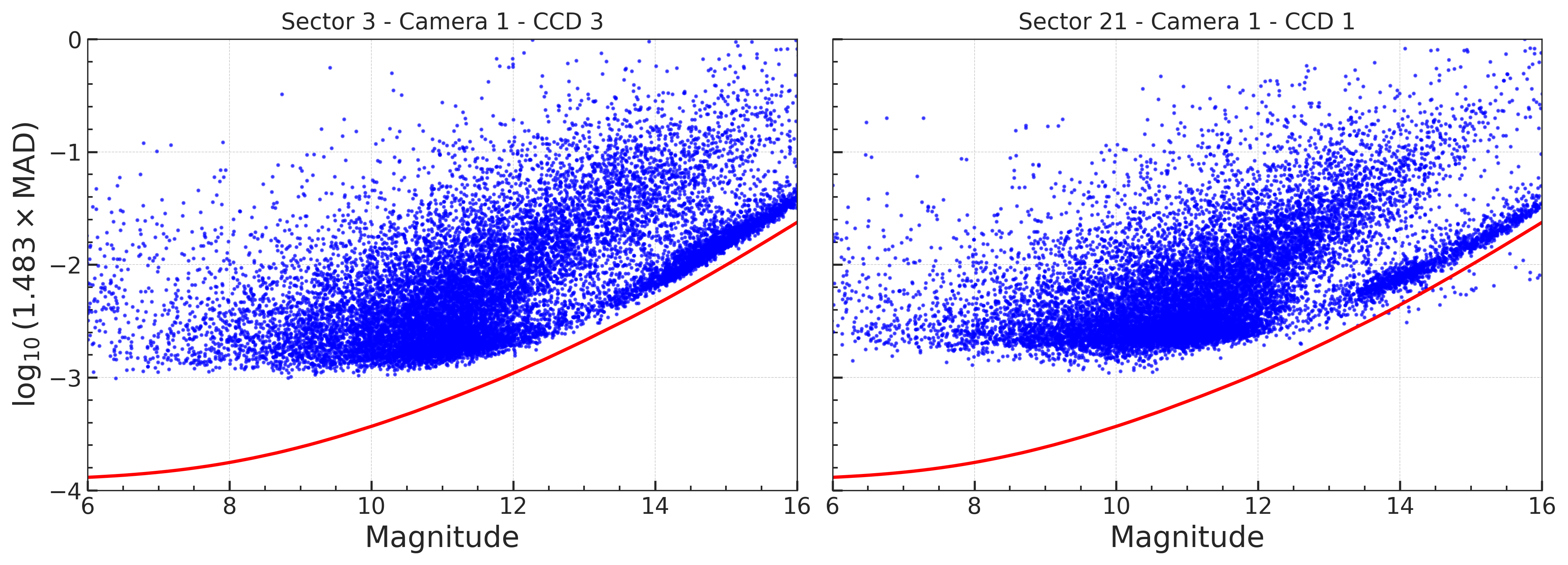}
    \caption{ Samples of RMS (calculated as 1.483 $\times$ MAD) against magnitude plots of the light curve obtained from the CCD 3 Camera 1 in sector 3 (left) and the CCD 1 Camera 1 of sector 21 (right) compared to the prelaunch noise-level expectation from \citet{Sullivan_2015} (solid red lines) scaled to 30-min FFIs cadence.}
    \label{Fig_magnitude_RMS}
\end{figure*}

\subsection{Signal and noise metrics}\label{sub_sec:noise_metric}

To assess the photometric quality of our pipeline, we computed three key metrics of the median-normalized light curves: the point-to-point median-differential variability (PTP-MDV), the root mean square (RMS) scatter, and the signal-to-noise ratio (S/N), defined as the ratio of RMS to PTP-MDV. These calculations were performed for each light curve within a given sector-camera-ccd combination. We adopted robust statistics based on the median to reduce the influence of outliers. 

The PTP-MDV, which serves as the estimate of the noise metric at the 30-minute observation cadence, is calculated as
\begin{equation} \label{eqn:MDV}
\text{MDV} = \text{median}\left(|F_{i+1} - F_{i}|\right) \, ,
\end{equation}
where $F_i$ and $F_{i+1}$ represent two subsequent median-normalized flux values at a given epoch for each source.

The RMS for each median-normalized light curve was estimated using the median absolute deviation (MAD), scaled by a factor of 1.483. This scaling arises because, for a Gaussian distribution, the MAD is related to the standard deviation by

\begin{equation}
\sigma = \frac{\mathrm{MAD}}{\Phi^{-1}(0.75)} \approx 1.483 \times \mathrm{MAD},
\end{equation}
where $\Phi^{-1}(0.75) \approx 0.6745$ is the 75th percentile of the standard normal distribution \citep{Hampel1974, RousseeuwCroux1993}. Thus, the scaled MAD provides a robust estimate of the standard deviation for large data samples, even if the data distribution deviates from the Gaussian distribution at the tails.
 
In Fig.~\ref{Fig_magnitude_RMS}, we present sample RMS–magnitude plots from both hemispheres. The left panel shows CCD~3 of Camera~1 in sector~3, representing the southern hemisphere, while the right panel shows CCD~1 of Camera~1 in sector~21 from the northern hemisphere. The prelaunch noise model of \citet{Sullivan_2015} is plotted in both panels with a solid line after being converted from its original units of ppm\,hr$^{1/2}$ to the 1800~s (0.5~h) exposure cadence of TESS FFIs. We assume that the floor of the RMS-magnitude distribution is determined by point sources whose light curves are dominated by noise rather than by signal. Typically, it shows two regimes: a noise floor of $10^{-3}$ for point sources brighter than $T_{\rm mag}\sim11$ and a gradual increase toward fainter magnitudes consistent with the predicted photon-noise limit for variable point sources. The former is probably related to uncorrected systematics, shared by all sources, while the latter originates from our deliberate choice of producing light curves only for variable sources. We remind that the sample of point sources that were selected for light curve extraction was deliberately biased against non-variable stars. This is why, compared with photometric surveys that extract light curves for all point sources in the FOV, the fraction of point sources close to the floor of the RMS–magnitude distribution in Fig.~\ref{Fig_magnitude_RMS} is relatively small. The discussion of whether, in light of the catalog objectives, such increased noise levels are acceptable, is deferred to Section \ref{sec:discussion}.

\subsection{Blending}\label{sub_sect:blending}
The DIA effectively subtracts non-varying sources from the residual images down to the photon noise levels and retains residual flux for variable objects in relation to the photometric master frame flux value. However, the method still encounters limitations when multiple nearby variable sources are present. Light dilution persists in the residual images in such cases, leading to blended light curves with bright sources affecting and dominating the flux of the dim ones. This becomes apparent in TESS data due to its relatively low angular resolution. Some sources show contamination from neighboring variable sources, especially in regions with high stellar density.

To guide our catalog users in identifying potential blending, we assess each light curve for possible contamination from nearby sources. Specifically, we search for sources within the aperture diameter, 6 pixels in both x and y coordinates—using the \texttt{cKDTree} function from the \texttt{astropy.spatial} module. The resulting table assigns a \texttt{group} identification number to each set of potentially blended sources within a given CCD. On average, each CCD contains about 7,400 groups, with a mean of two stars per group. It is important to note that this threshold is conservative and does not definitively indicate blending; users are encouraged to verify potential contamination in their light curves of interest.

\subsection{False positives}\label{sub_sect:false}

   \begin{figure}
   \centering
   \includegraphics[width=0.49\textwidth, keepaspectratio]{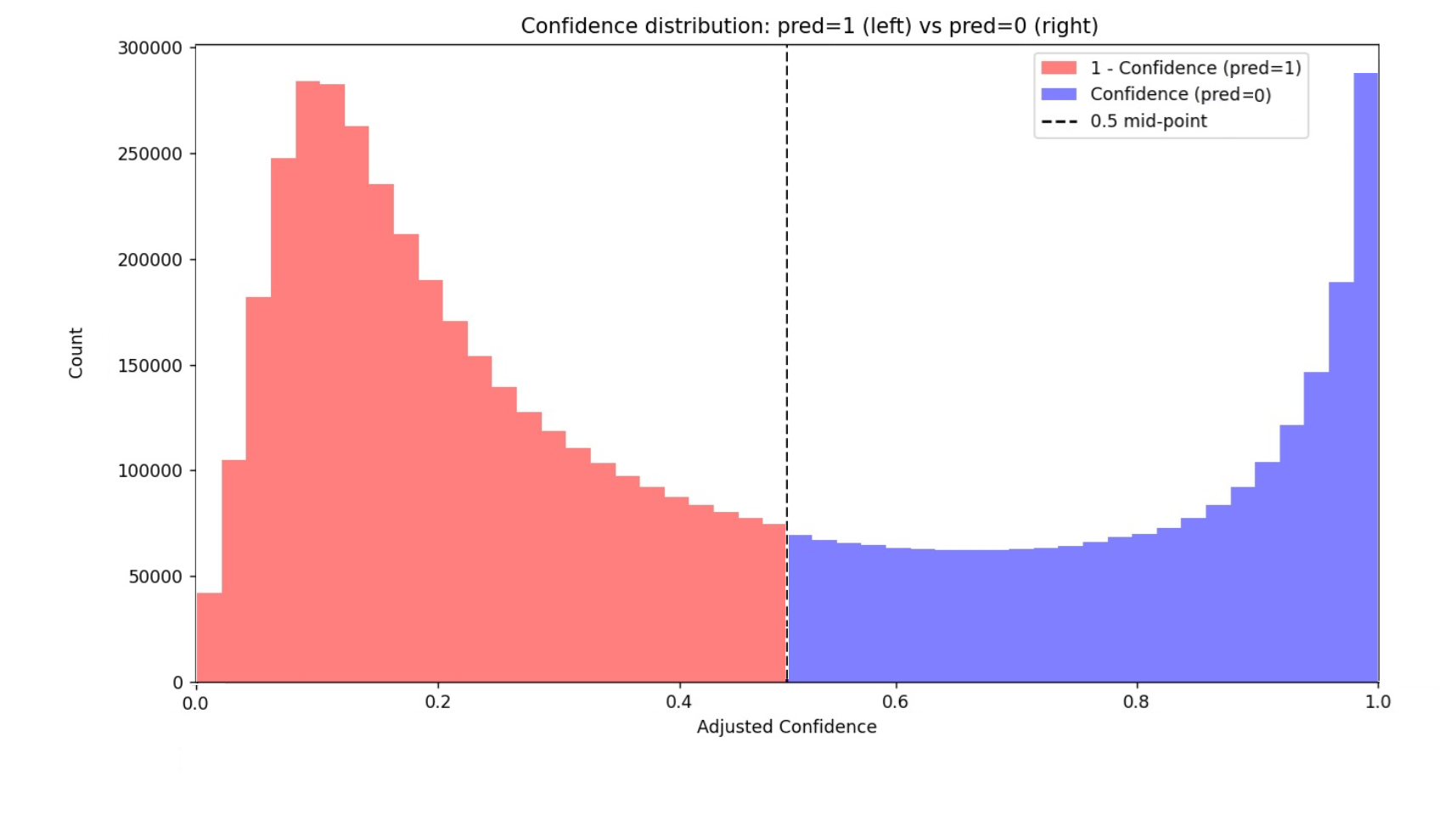}
   \caption{Overall distribution of the confidence for both classes in the binary CNN classification. The confidence distribution of the ``systematic class'' (``pred=1,'' red) was inverted around the midpoint ($0.5$).}
              \label{Fig_FP}%
    \end{figure}

DIA is known to be prone to spurious and false detections (e.g., \citealt{Wozniak2000, Wright2015}), which has motivated the development of various statistical approaches to improve reliability \citep[e.g.,][]{Sokolovsky_2017}. This issue is particularly evident in TESS data, where orientation and distance variations of the Earth and Moon relative to the camera boresight introduce scattered light into the field of view. These changes modulate the background flux over time—most noticeably near the end of each orbit when the Earth rises above the sunshade—and can mimic astrophysical variability, leading to spurious detections of candidate variable sources.\footnote{Patterns in the background behavior are documented in the data release notes for each sector at \url{https://archive.stsci.edu/tess/tess_drn.html}}

Machine-learning methods have been increasingly employed in light curve analysis (e.g., \citep{Cui_2024, Akhmetali2024, Yu_2021, Hinners_2018}) for asks such as classification, prediction, and anomaly detection. Building on these advances, we developed a supervised deep-learning classifier to distinguish genuine astrophysical variability from signals dominated by instrumental effects. Each light curve in the catalog was evaluated by this model and assigned a label, either predominantly systematic or astrophysical, together with a confidence score indicating the reliability of the classification (see Table~\ref{tab:photometry_columns}).

After evaluating the performance of several pretrained Convolutional Neural Network (CNN) architectures, we opted to develop a custom CNN tailored to the light curves in our catalog. Full details of the network architecture and implementation are available in our public GitHub repository\footnote{\url{https://github.com/tesslc2025-star/TESS-LC-Prim-Classification}}

The training dataset consisted of 6,656 light curves, randomly chosen from all sector-camera-ccd combinations. The corresponding FITS files were normalized and converted from their numerical time-series representation into PNG image format. Each light curve was manually inspected and classified as mostly containing systematic instrumental noise or containing true-variability signal, thereby providing the labeled dataset required to train and evaluate our CNN classifier within a supervised machine learning framework. The resulting dataset was well-balanced, comprising 3,418 light curves labeled as likely systematic (`1') and 3,238 as likely containing true variability (`0').

The classification model was evaluated on a held-out test set of 1,332 light curves, achieving an overall accuracy of 77.5\% (Table~\ref{tab:cnn_eval}). The detailed class-level metrics, including precision, recall, and F1-scores for both classes, are reported in the table. Macro-averaged and weighted-averaged metrics further confirm the effectiveness of generalization without significant bias toward either class.

For each classification, the model’s confidence level in the selected class is also provided   (Table~\ref{tab:photometry_columns}). Although a high confidence does not necessarily correspond to a high probability of a correct classification, it is correlated with it, especially considering the network’s high accuracy.

Overall, $62.2$\,\% of the light curves were classified as systematic (`1') and $37.8$\,\% of the light curves were classified as `true variability' (`0'). Figure \ref{Fig_FP} shows the overall distribution of the confidence for both classes in the binary CNN classification. To include both classes in one plot, the confidence distribution of the `systematic' class (`1') was inverted around the midpoint ($0.5$). The CNN classification clearly separates the two classes, albeit not perfectly. Interestingly, the confidence mode for the class 'true variability' (`0') is close to $1.0$, whereas for the `systematic' class (`1') it is closer to $0.9$. Despite its imperfections, we hope that catalog users will find this binary classification and the associated confidence values useful in determining the choice of light curve samples they wish to investigate. In Section \ref{sec:future_plans}, we outline possible future improvements of the method.

\begin{table}[h!]
\centering
\caption{Final evaluation metrics of the CNN classifier on the held-out test set.}
\label{tab:cnn_eval}
\begin{tabular}{lccc}
\toprule
Class / Average & Precision (\%) & Recall (\%) & F1-score \\
\midrule
Signal (`0') & 81.4 & 67.7 & 0.74 \\
Systematic (`1') & 74.9 & 86.2 & 0.80 \\
\midrule
Macro Avg    & 78.2 & 76.9 & 0.77 \\
Weighted Avg & 78.3 & 77.5 & 0.77 \\
\midrule
Overall Accuracy & \multicolumn{3}{c}{77.5} \\
\bottomrule
\end{tabular}
\end{table}

\begin{figure}
    \centering
    \includegraphics[width=0.95\linewidth, height=18cm]{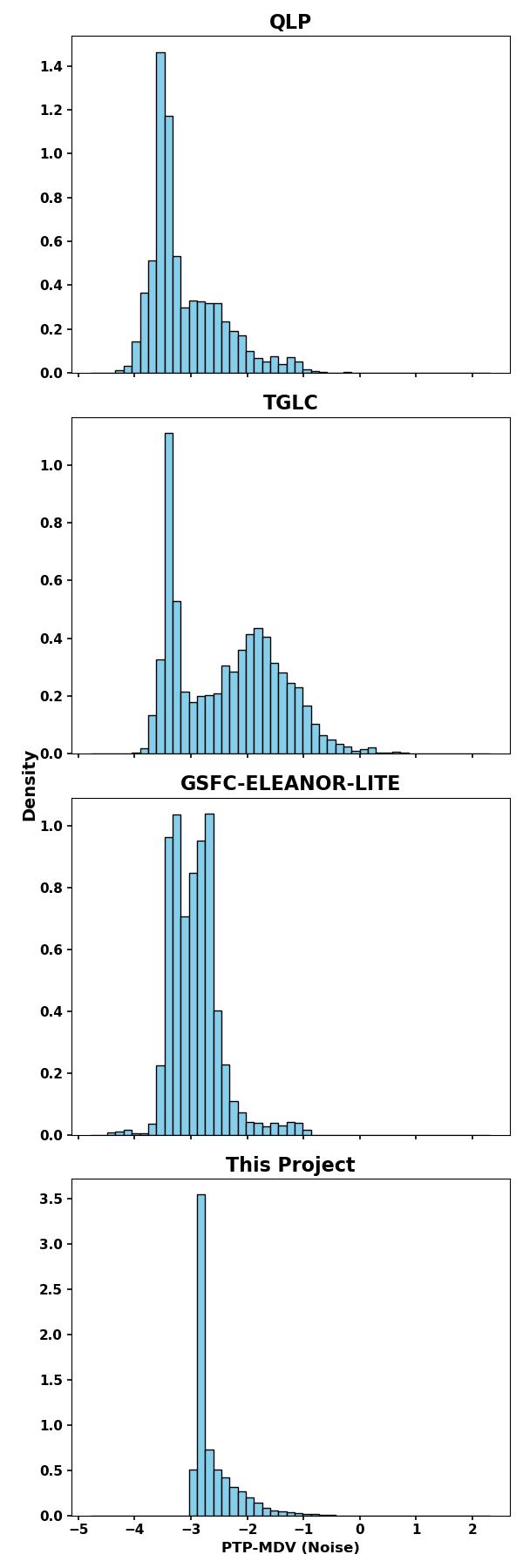}
    \caption{Distribution of noise levels (measured as the $\log_{10}$, PTP-MDV for light curves extracted from CCD 1, Camera 2, sector 6, comparing three TESS FFI light curve projects with this project.}
    \label{fig_noise_comparison}
\end{figure}

\vspace{1em}

\begin{figure*}
    \centering

    \begin{subfigure}{0.85\linewidth}
        \centering
        \includegraphics[width=\linewidth]{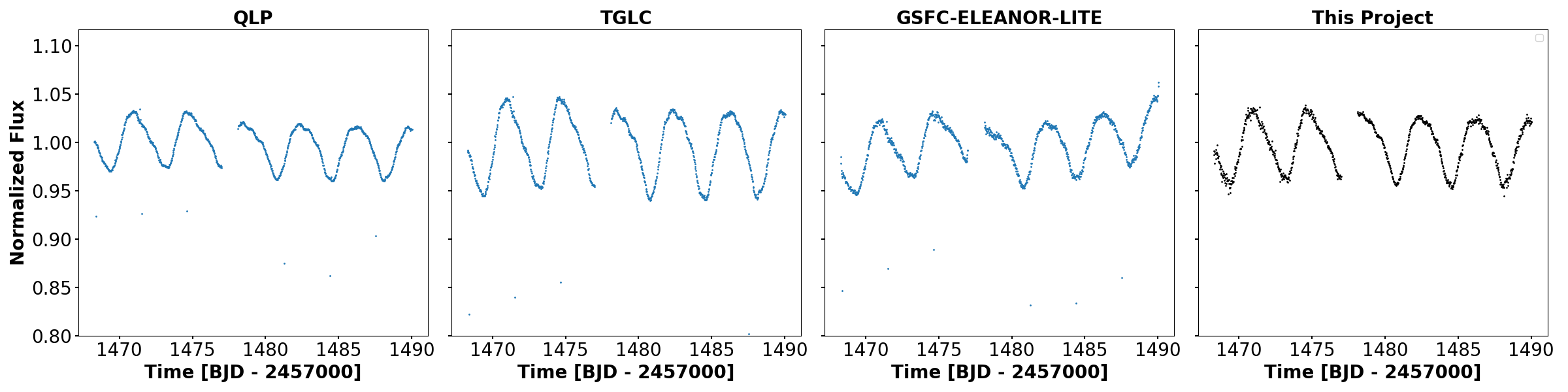}
        \caption{light curve of a periodic variable, TIC 32815829, T$_{\text{mag}}$= 11.84}
        \label{subfig:periodic_variable}
    \end{subfigure}

    \vspace{0.5em}

    \begin{subfigure}{0.85\linewidth}
        \centering
        \includegraphics[width=\linewidth]{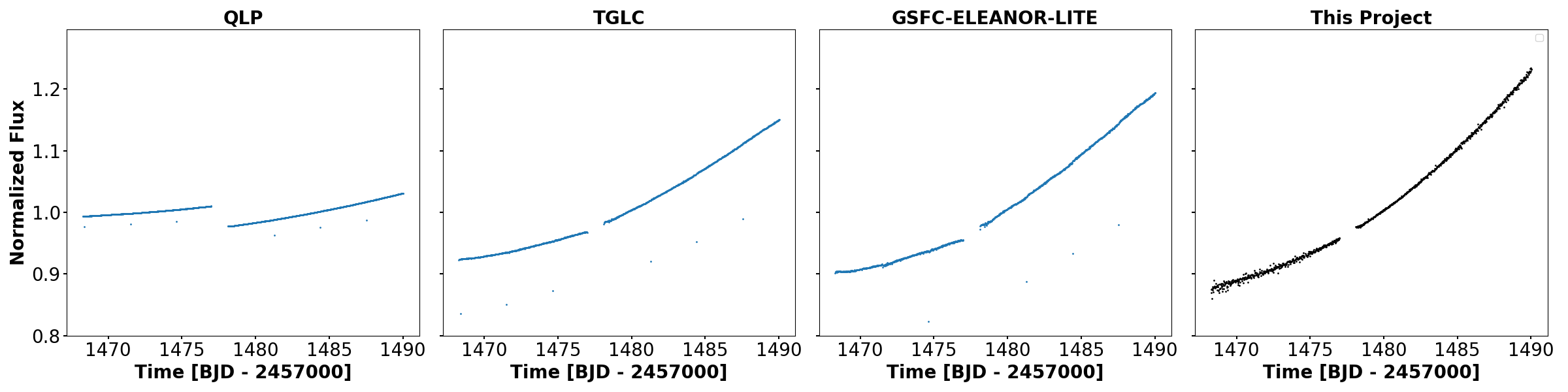}
        \caption{light curve of a Mira variable star TIC 31357861, $T_{\text{mag}}$= 9.82}
        \label{fig:long_period_variable}
    \end{subfigure}

    \vspace{0.5em}

    \begin{subfigure}{0.85\linewidth}
        \centering
        \includegraphics[width=\linewidth]{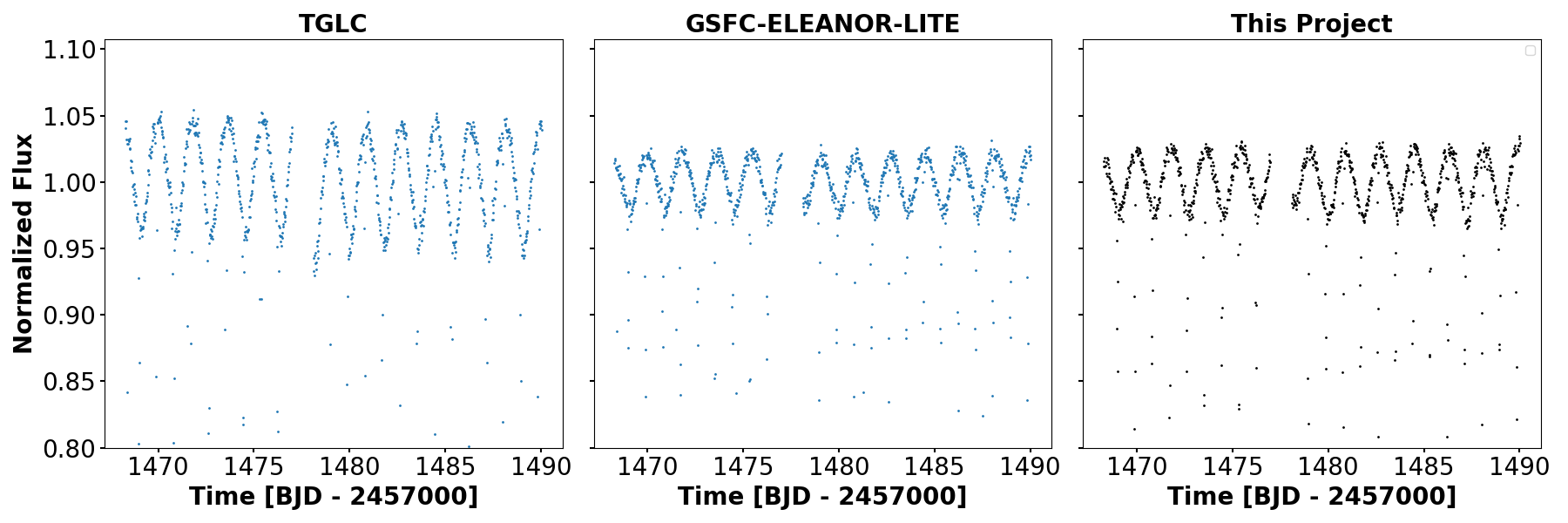}
        \caption{light curve of an eclipsing binary, TIC 32539020, $T_{\text{mag}}$= 14.10}
        \label{fig:sub3}
    \end{subfigure}

    \caption{Comparison of our light curves for various types of variable stars,  with different magnitude ranges, with those from QLP, TGLC, and GSFC-ELEANOR-LITE. All sources are from CCD 1 of Camera 2 in sector 6.}

    \label{fig:comparison_plots}
\end{figure*}

\begin{table*}[tbp]
    \centering
    \caption{Description of the main data columns in the photometric dataset.}
    \label{tab:photometry_data}
    \begin{tabular}{lllll}
        \toprule
        Keyword & Column Name & Data Type & Unit & Description \\
        \midrule
        TIME         & Time               & Double  & JD & Observation time \\
        d\_FLUX      & Differential Flux  & Float   & e$^-$/s    & Flux change relative to reference \\
        d\_FLUX\_ERR & Differential Flux Error & Float   & e$^-$/s    & Uncertainty in differential flux \\
        SAP\_QUALITY & Quality Flag       & Integer & -         & Data quality flag \\
        FLUX         & Flux               & Float   & e$^-$/s    & Absolute flux measurement \\
        FLUX\_ERR    & Flux Error         & Float   & e$^-$/s    & Uncertainty in flux \\
        \bottomrule
    \end{tabular}
\end{table*}

\section{catalog description} \label{sec:productdesc}

The primary data products in this project are extracted light curves stored in the standard binary FITS format. Each light curve corresponds to a specific CCD within a given camera and sector.

The naming convention for the products on MAST is:
\begin{center}\footnotesize
\texttt{hlsp\_tequila\_tess\_ffi\_s\{sector\}-cam\{camera\}-ccd\{ccd\}\\
\-x\{pixel\_index\}-y\{pixel\_index\}\_tess\_\\
v<version>\_llc.fits}.
\end{center}
Here, \texttt{sector}, \texttt{camera}, and \texttt{ccd} indicate the TESS sector (zero-padded to four digits), camera, and CCD of the observation, respectively.   
\texttt{x\{pixel\_index\}} and \texttt{y\{pixel\_index\}} denote the four-digit zero-padded pixel coordinates of the target centroid. 

For example, the file \texttt{hlsp\_tequila\_tess\_ffi\_s0008-\\
cam2-ccd2-x0005-y0197\_tess\_v1\_llc.fits} corresponds to a target observed in Sector 8, Camera 2, CCD 2, and centroid pixel coordinates $(x, y) = (5, 197)$. The current version is version 1 of the TEQUILA database. In future enhancements of the pipeline and data, the version will be incremented.

Metadata is embedded within the primary header (see Appendix \ref{app:lc_info_tab} for an example). As described in Table \ref{tab:photometry_data}, the primary data columns include time, differential flux, differential flux error, the SAP\_QAULITY flag, which defaults to 0, flux, and flux error. Additional data products include photometric master calibration frames and the summary tables listing all detected sources with key attributes (see Table~\ref{tab:photometry_columns} for further descriptions).

\section{Discussion}\label{sec:discussion}

In this work, we conducted an untargeted search and generated light curves for variable point sources observed during the TESS primary mission using differential image analysis. Our approach produced over six million candidate light curves encompassing variable stars, transients, and Solar System objects. 

We achieved noise levels, as estimated by the median point-to-point variability, on the order of \(10^{-3}\) for bright stars, with fainter objects naturally exhibiting higher noise. Despite our careful estimation methods, effects from scattered light, saturation, and blending persist in portions of the dataset. To aid users in assessing data quality, we provide a table summarizing key information for each light curve, including group identification numbers for potential blending, a flag that indicates possible relation to a known SSO, and a CNN-based confidence level of whether the light curve was created due to systematic instrumental noise or due to true variability. 

This catalog is intended as an exploratory tool for investigating variable objects in TESS. To compare our results with other TESS FFI light curves, particularly the stellar variables in the projects listed in Table~\ref{tab:ffiprojects}, we queried the coordinates of our sources to obtain their TIC identification numbers and cross-matched them with relevant projects to retrieve their light curves from MAST. Specifically, we compared our light curves with the raw, undetrended light curves version from MIT-QLP \citep{2020Huang}, TGLC \citep{2023Han}, and GSFC-ELEANOR-LITE \citep{2022Powel}. We retrieved approximately 4,600 matched light curves from 18,182 extracted point sources in CCD 1 of Camera 2 in sector 6, using a 21-arcsecond matching radius.

Figure \ref{fig_noise_comparison} presents a comparative distribution of the PTP-MDV noise metric for QLP, TGLC, GSFC-ELEANOR-LITE, and our project. The distributions for TGLC and GSFC-ELEANOR-LITE show evidence of bimodality, with a primary population of high-quality light curves (mode near log$_{10}$(PTP-MDV) $\sim -3.5$) and a secondary population of noisier data. In contrast, the QLP distribution appears largely unimodal, with a peak at lower PTP-MDV values. Our project also shows a unimodal distribution, centered at log$_{10}$(PTP-MDV) $\sim -2.8$. Although our median noise level is higher than the leading mode in other projects, we note that our pipeline was rather optimized for discovery than for S/N. 
Future improvements may reduce our noise level further, but even at this stage, our method produces a large dataset of potentially variable sources across a wide magnitude range. 

In terms of photometric precision, we evaluated the S/N, as described in Sec. \ref{sub_sec:noise_metric}, for the representative subset of the retrieved variable stars across these projects. Out of the matches, we find that 2,087 out of 2,310 (90.4\%) from QLP, 4,328 out of 4,564 (94.8\%) from TGLC, and 4,372 out of 4,550 (96.1\% ) from GSFC-ELEANOR-LITE exhibit higher S/N than ours, which reflects the targeted source extraction design of these pipelines. Our light curves are noisier in 90–96\% of the overlapping cases, but noise minimization is not the primary aim of our pipeline. Instead, our goal is to deliver a uniform and flexible extraction approach that ensures completeness across all variable types in the TESS FFIs. Once a target of interest is found, the S/N can be improved by applying a tailored photometry, such as those listed in Table \ref{tab:tess_tools}, for a customized extraction technique.

Representative samples are illustrated in Figure~\ref{fig:comparison_plots} for different magnitude ranges. The upper panel displays a periodic variable star, TIC 32815829, with a $T_{\text{mag}}$ of 11.84. The middle panel shows a Mira (long-period) variable star, TIC 31357861, with a $T_{\text{mag}}$ of 9.82, and the bottom panel features an eclipsing binary, TIC 32539020, with a $T_{\text{mag}}$ of 14.10—above the QLP magnitude threshold—where our light curve performs comparably well.

\begin{figure}[ht]
    \centering
    \includegraphics[width=\linewidth, height=0.2\textheight]{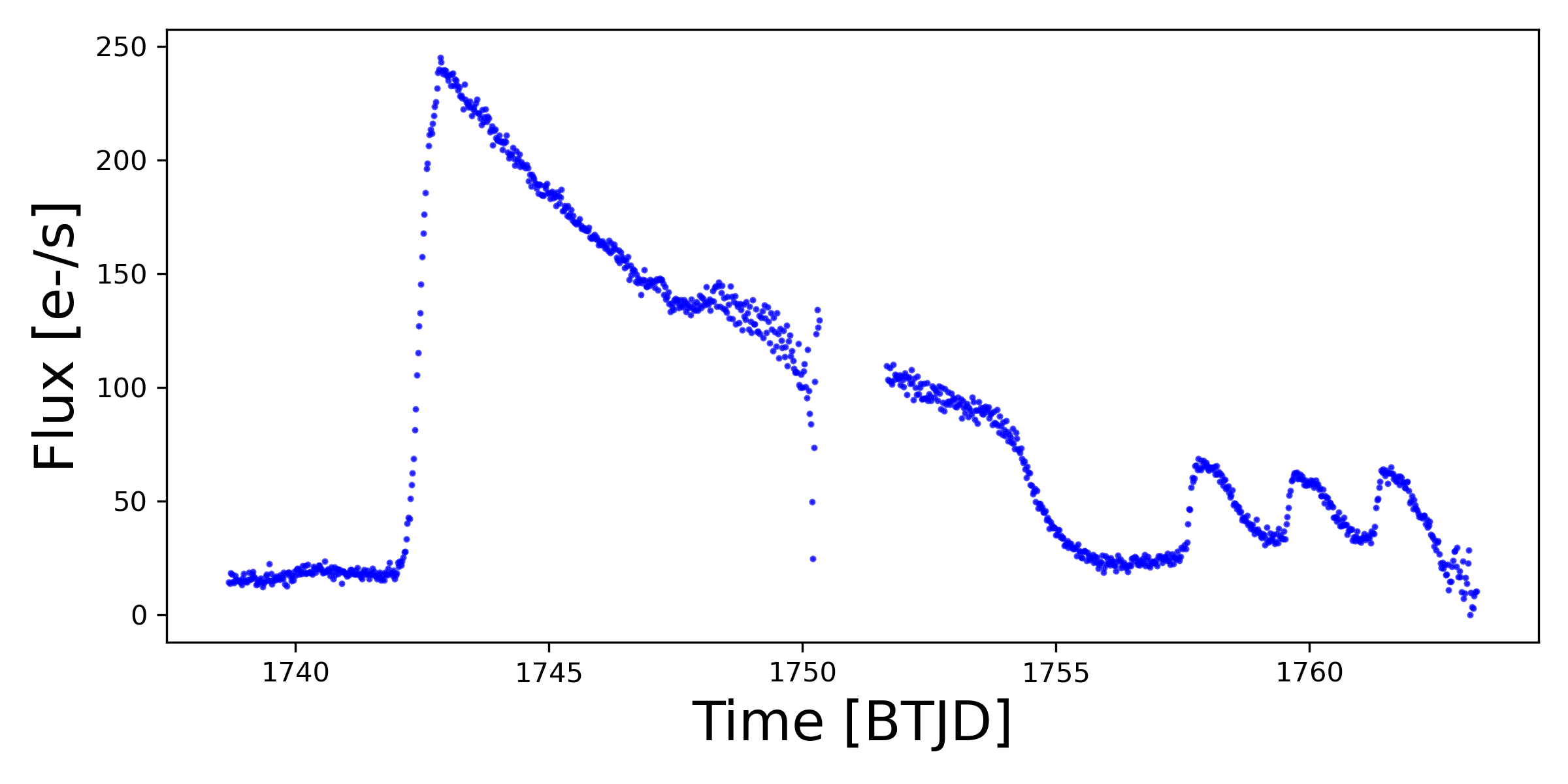}
    \caption{Light curve of the transient ASASSN-15nl extracted with TEQUILA. 
        At quiescence, the source is faint ($T_{\rm mag} \sim 19.5$), but the superoutburst enabled its recovery, illustrating the pipeline’s ability to detect transients otherwise missed by other catalogs.}

    \label{fig:assasn15nl}
\end{figure}

Another important contribution of our catalog is the inclusion of light curves for transient events, which are largely absent from other TESS FFI catalogs. In Fig.~\ref{fig:assasn15nl}, we show the light curve of TIC~1000887074 (Gaia DR2 1484498477416555904), also known as ASASSN-15nl \citep{Kato_2016}, a dwarf nova that underwent a superoutburst during sector~16 observations. At quiescence, the object has a TESS magnitude of 19.46, below the nominal limits of existing TESS FFI projects listed in Table~\ref{tab:ffiprojects}. However, the transient brightening during the outburst brought it into the detectable range, allowing TEQUILA to extract its light curve. This example demonstrates that our approach enables recovery of transients and variables that fall outside the magnitude limits of other pipelines in their quiescent states, thereby expanding the accessible discovery space.

This catalog represents the latest addition to the TESS HLSP light curve collection, focusing specifically on variable point sources. It primarily includes variable stars, small Solar System objects (SSOs), and transients. This focus will facilitate further scientific exploration using the TESS FFI light curves and enable a more comprehensive study of time-domain astronomy. For instance, in stellar classification, most classifications conducted to date have primarily relied on the 2-minute cadence data \citep[e.g.,][]{2025Wang, Gao_2025}. With this catalog, we aim to classify all sources, potentially including those not covered by existing catalogs. Additionally, the catalog provides an opportunity to conduct a broader study on stellar statistics such as rotation periods in fast-rotating stars, complementing existing research \citep[e.g.,][]{Colman_2024}.

\section{Summary and future plans}\label{sec:future_plans}

In the near term, we plan to implement corrections for systematic noise to further improve the data quality and to extend our work to include data from the TESS extended missions. At this stage, we have not yet attempted correcting for the scattered light beyond the simplistic method described in Section \ref{sec:method}. Neither have we applied detrending corrections, such as sysrem \citep{2005Tamuz}, a common practice in similar projects. Before extracting FFI-based light curves from the extended TESS mission, we will attempt to remove systematic instrumental noise at both image (preprocessing) and light curve (postprocessing) stages. In addition, we will optimize the photometry to minimize the gap between the observed and predicted noise floors at the photon-limited side of the RMS-magnitude diagram (see Figure \ref{fig_noise_comparison})

As a next step, we will explore a range of light curve classification techniques. Alongside traditional methods used in time-series analysis, we will investigate both supervised and unsupervised machine-learning approaches to evaluate their effectiveness in identifying scientifically interesting light curves. Particular emphasis will be placed on the detection and characterization of periodic signals, such as those arising from variable stars, eclipsing binaries, and potential exoplanet hosts. By comparing the performance of different approaches, we aim to establish a framework that efficiently prioritizes light curves for follow-up analysis.

In parallel, we are developing a more robust pipeline for the detection and identification of Solar System Objects (SSOs). This will enable more reliable flagging of sources and light curves associated with known SSOs by accounting for their unique variability patterns. Using forced photometry along predicted trajectories, we will also generate dedicated SSO light curves, opening the door to additional science cases such as the study of asteroid rotation periods.

In summary, this work establishes the foundation for a database designed to support a broad range of astrophysical investigations. By prioritizing an untargeted, raw-data approach, we maximize the catalog’s potential for discovery while leaving room for user-driven customization and refinement. As TESS continues to deliver new data, we anticipate that our methods and infrastructure will evolve in step, contributing to a deeper understanding of the dynamic universe.

\section*{Data availability}
The light-curve products are accessible on MAST as a high-level science product via \url{https://doi.org/10.17909/pj9v-7t21}\footnote{\url{https://archive.stsci.edu/hlsp/tequila/}}
. The catalog data are available in electronic form at the CDS, either via anonymous FTP to \texttt{cdsarc.u-strasbg.fr} (130.79.128.5) or through the web interface at \url{http://cdsweb.u-strasbg.fr/cgi-bin/qcat?J/A+A/}
.

\begin{acknowledgements}
The authors express gratitude to Avi Shporer for his feedback on this paper. BBO acknowledges the Kavli Institute for Astrophysics and Space Research at the Massachusetts Institute of Technology for hosting him as a visiting research student during part of this work. BBO and YZ acknowledge the Israel Data Science and AI Initiative (IDSAI) 2023 grant for the purchase of cloud computing resources. This research was partially funded by the Israel Ministry of Innovation, Science, and Technology through Grant number 0008107, and by the Israel Science Foundation through grant No. 1404/22. This project includes data collected with the TESS mission from the MAST data archive at the Space Telescope Science Institute (STScI). The NASA Explorer Program provides funding for the TESS mission. STScI is operated by the Association of Universities for Research in Astronomy, Inc., under NASA contract NAS 5–26555.
\end{acknowledgements}

\bibliographystyle{aa}
\bibliography{references}

\begin{appendix}
\onecolumn
\section{Light curve information tables}\label{app:lc_info_tab}

The light curve FITS files follow the standard \texttt{Flexible Image Transport System (FITS)} convention. Each file consists of two primary components, or Header Data Units (HDUs): 
\begin{itemize}
    \item HDU 0 (Primary Header): Contains global metadata describing the observation context, instrument setup, and object coordinates.
    \item HDU 1 (Extension Header): Contains tabulated time-series photometric data (light curve) and its associated uncertainties, along with metadata describing the data format.
\end{itemize}

The primary header of each catalog file includes all essential details about the extracted point source. Table~\ref{tab:primary_header} presents a sample header of a light curve from CCD~1, Camera~1 of sector~4.

\begin{table*}[htbp]
\centering
\caption{Sample FITS primary header information (sector~4, Camera~1, CCD~1)}
\label{tab:primary_header}
\begin{tabular}{|l|l|l|l|}
\hline
Keyword & Example/Default Value & Data Type & Comment \\
\hline
SIMPLE   & T                     & bool   & Conforms to FITS standards \\
BITPIX   & 8                     & int    & Array data type \\
NAXIS    & 0                     & int    & Number of array dimensions \\
EXTEND   & T                     & bool   & File contains extensions \\
NEXTEND  & 2                     & int    & Number of standard extensions \\
EXTNAME  & PRIMARY               & str    & Name of extension \\
EXTVER   & 1                     & int    & Extension version number (not format version) \\
ORIGIN   & AU/AGASS              & str    & Institution responsible for file \\
DATE     & 2024-06-09            & str    & File creation date \\
CREATOR  & lightkurve.LightCurve.to\_fits() & str & Pipeline job and program used \\
TELESCOP & TESS                  & str    & Telescope \\
INSTRUME & TESS Photometer       & str    & Detector type \\
PROCVER  & 2.0.11                & str    & Processing version \\
CCD      & 4                     & int    & CCD number used for the observation \\ 
CAMERA   & 1                     & int    & Camera number used for the observation \\ 
SECTOR   & 6                     & int    & TESS sector number \\
BUNIT    & e$^-$/s               & str    & Units of flux \\
HLSPNAME & TESS Quick-Look and Light curve Analysis & str & High-Level Science Product name \\
HLSPID   & TEQUILA               & str    & HLSP identifier \\
HLSPLEAD & Bisi Ogunwale         & str    & HLSP lead \\
HLSP\_PI & Lev Tal-Or            & str    & Principal Investigator \\
HLSPVER  & v1                    & str    & HLSP version \\
DOI      & 10.17909/pj9v-7t21    & str    & Digital Object Identifier \\
TIMESYS  & UTC                   & str    & Time system used \\
HLSPTARG & s0004-cam1-ccd1-x0003-y0613 & str & Target ID for HLSP \\
TSTART   & 1468.277              & float  & Observation start time [BTJD] \\
TSTOP    & 1490.0266             & float  & Observation stop time [BTJD] \\
MJD-BEG  & 58467.777             & float  & Start of observation [MJD] \\
MJD-END  & 58489.5266            & float  & End of observation [MJD] \\
DATE-BEG & 2018-12-15T18:38:50.067 & str  & Start of observation [ISO-8601] \\
DATE-END & 2019-01-06T12:38:21.698 & str  & End of observation [ISO-8601] \\
TELAPSE  & 21.75                 & float  & Time elapsed (TSTOP - TSTART) [d] \\
\hline
\end{tabular}
\end{table*}

\vspace{1em}

The secondary (extension) header contains metadata describing the structure and format of the light curve table data. Table~\ref{tab:extension_header} lists a representative example from sector~4, Camera~1, CCD~1.

\begin{table*}[htbp]
\centering
\caption{Sample secondary header (extension HDU~1) information}
\label{tab:extension_header}
\begin{tabular}{|l|l|l|l|}
\hline
Keyword & Example/Default Value & Data Type & Comment \\
\hline
XTENSION & BINTABLE & str & Binary table extension \\
BITPIX   & 8 & int & Array data type \\
NAXIS    & 2 & int & Number of array dimensions \\
NAXIS1   & 28 & int & Length of dimension~1 \\
NAXIS2   & 959 & int & Length of dimension~2 \\
PCOUNT   & 0 & int & Number of group parameters \\
GCOUNT   & 1 & int & Number of groups \\
TFIELDS  & 6 & int & Number of table fields \\
EXTNAME  & LIGHT CURVE & str & Name of extension \\ 
RA\_TARG & 88.8872 & float & Target right ascension [deg] \\
DEC\_TARG & 13.8036 & float & Target declination [deg] \\
RADESYS & ICRS & str & Reference system \\
FILTER  & TESS & str & Photometric filter \\
CADENCE & 1800 & int & Time between data points [s] \\
TTYPE1  & TIME & str & Column name for time data \\
TFORM1  & D & str & Data type for TIME (float64) \\
TUNIT1  & jd & str & Unit of TIME (Julian days) \\
TTYPE2  & d\_FLUX & str & Detrended flux column \\
TFORM2  & E & str & Data type for d\_FLUX (float32) \\
TUNIT2  & e$^-$/s & str & Unit of d\_FLUX \\
TTYPE3  & d\_FLUX\_ERR & str & Error in detrended flux \\
TFORM3  & E & str & Data type for d\_FLUX\_ERR (float32) \\
TUNIT3  & e$^-$/s & str & Unit of d\_FLUX\_ERR \\
TTYPE4  & SAP\_QUALITY & str & Quality flag column \\
TFORM4  & J & str & Data type for SAP\_QUALITY (int32) \\
TTYPE5  & FLUX & str & Raw flux column \\
TFORM5  & E & str & Data type for FLUX (float32) \\
TTYPE6  & FLUX\_ERR & str & Error in raw flux \\
TFORM6  & E & str & Data type for FLUX\_ERR (float32) \\
\hline
\end{tabular}
\end{table*}

\vspace{1em}

Each CCD of a given camera in a sector is accompanied by a summary table containing all extracted light curves. The naming convention of each table follows the format:

\begin{center}
\footnotesize
\texttt{s\{sector\}-\{camera\}-\{CCD\}\_master\_frame\_photometry\_.csv}
\end{center}

Table~\ref{tab:photometry_columns} provides a description of the summary table headers, along with the meaning of each column.

\begin{table*}[htbp]
\centering
\caption{Description of the \texttt{master\_frame\_photometry} file data dolumns for e1ach CCD}
\label{tab:photometry_columns}
\begin{tabular}{@{}l p{12cm}@{}}
\toprule
Key & Description \\
\midrule
\texttt{ID} & Unique identifier for each source. \\
\texttt{Filename} & Name of the FITS file corresponding to the source light curve. \\
\texttt{x-center} (px) & X-coordinate of the source center in pixels on the FFI. \\
\texttt{y-center} (px) & Y-coordinate of the source center in pixels on the FFI. \\
\texttt{Aperture Sum} & Total flux measured within the photometric aperture for the source, used as the reference flux in constructing its light curve. \\
\texttt{RA\_OBJ} (J2000) & Right Ascension of the source (J2000). \\
\texttt{DEC\_OBJ} (J2000) & Declination of the source (J2000). \\
\texttt{Flux} & Reference flux after background subtraction. \\
\texttt{Flux Error} & Estimated uncertainty in the reference flux measurement. \\
\texttt{Magnitude} & Estimated TESS magnitude (see Section~\ref{sub-sec:flux_cal}). \\
\texttt{Magnitude Error} & Uncertainty in the magnitude estimation. \\
\texttt{Noise\_Level} & Estimated noise level in the measurement (PTP–MDV; see Section~\ref{sub_sec:noise_metric}). \\
\texttt{RMS} & Root Mean Square ($1.483 \times$ MAD) of the flux measurement. \\
\texttt{Group} & Identifier for potential blending group sources (see Section~\ref{sub_sect:blending}). \\
\texttt{SNR} & Signal-to-Noise Ratio (\texttt{RMS}/\texttt{Noise\_Level}). \\
\texttt{sso\_contam} & Boolean flag indicating possible `contamination' by a known SSO (see Section~\ref{sec:ssos}). \\
\texttt{is\_systematic} & Boolean flag for likely systematic noise (see Section~\ref{sub_sect:false}). \\
\texttt{confidence} & Confidence level of the binary classification (see Section~\ref{sub_sect:false}). \\
\bottomrule
\end{tabular}
\end{table*}

\end{appendix}

\end{document}